\documentclass[epj
]{svjour}

\usepackage{epsfig}
\usepackage{color}

\newcommand{\beq}{\begin{equation}}
\newcommand{\eeq}{\end{equation}}
\newcommand{\beqa}{\begin{eqnarray}}
\newcommand{\eeqa}{\end{eqnarray}}

\newcommand{\ket}[1]{|#1\rangle}
\newcommand{\bra}[1]{\langle #1|}

\newcommand{\nn}{{\nonumber}\\}
\def\ten{\otimes}
\def\VacBra{\bra 0}
\def\VacKet{\ket 0}
\def\VacExpect #1{\VacBra #1\VacKet}
\def\Id{{\rm 1\kern-.3em I}}

\begin{document}

\title{Electroweak form factors of non-strange baryons} 

\author{Dirk Merten \and Ulrich L\"oring \and Klaus Kretzschmar \and Bernard Metsch \and Herbert-R. Petry}

\institute{Institut f\"ur Theoretische Kernphysik, Nu{\ss}allee 14--16, D-53115 Bonn, Germany, \email{merten@itkp.uni-bonn.de}}

\date{\today}
\abstract{We compute electroweak form factors of the nucleon and
photon transition form factors of non-strange baryon resonances up to the third
resonance region in a model with instanton-induced interaction. The
calculation is based on the Bethe-Salpeter equation 
for three light constituent quarks and is fully relativistic
(U. L\"oring {\it et al.}, Eur. Phys. J. {\bf A10}, 309 (2001)). Static
nucleon properties and photon resonance couplings are in good
agreement with  experiment and the $Q^2$-behaviour  of the
experimentally known form factors up to large 
momentum transfer is accounted for. 
}

\maketitle


\renewcommand{\arraystretch}{1.25}

\section{Introduction}
The notion of constituent quarks has proven to be a most successful
concept for the interpretation of hadron resonances. We know, however,
that they can, at best, be quasiparticles which arise due to a
dynamical break-down of chiral symmetry in fundamental QCD. The
details of this process are  still obscure; we believe, therefore, that
it is of general interest to work out precisely at which energies
this concept tends to fail. As a first step it is therefore necessary
to construct a constituent quark model, which fits the mass spetrum
quantitatively  and to extend this model in energy and momentum
transfer as much as possible. For this reason the following issues are
most essential for our approach:

1. The model has to be relativistical covariant. This is
   obviously not the case in non-relativistic quark models, which
   dominated the research of the last decades \cite{Isgur1}. The need for
   covariance was clearly recognised in the past, and led to various
   relativizations ({\it e.g.} \cite{Isgur2}\cite{Glozman}). They include in particular the use
   of formally relativistic energies and a prescription for
   boosting non-relativistic wave functions. The importance of this
   last point was recently demonstrated again by the Graz group \cite{Graz}. In
   contrast to this we have used in our attempts directly the
   Bethe-Salpeter equation with instantaneous forces, which respects
   covariance right from the beginning. This model was used for light
   quark flavours with considerable success concerning the spectrum of
   mesons \cite{Koll} and baryons \cite{Uli1}-\cite{Uli3}. With only 7 model parameters (masses and couplings)
   an efficient and good description of the experimentally known baryon spectrum
   could be achieved describing both Regge trajectories and the
   fine structure in the mass splittings. 

2. Usual non-relativistic quark inter\-ac\-tions (string-like
   confinement and one-gluon-exchange fine structure interaction)
   have proven to lead to inconsistent spectroscopic results,
   {\it e.g.} too large spin-orbit couplings. In addition the known nucleon
   Regge trajectory is not correctly explained. In our model we keep the
   string-like confinement and solve the spin-orbit problem by an
   appropriate Dirac structure. We also replace the one-gluon exchange
   by 't~Hooft's instanton-induced interaction
   \cite{Hooft}\cite{Shifman}. This has the advantage that 
   the $U_A(1)$ anomaly of the meson spectrum is correctly treated and
   that the $\eta'$-problem is convincingly solved from the very
   beginning. In addition 't~Hooft's force is very similar in structure
   to the dynamics of a Nambu-Jona-Lasinio model and offers a natural 
   mechanism for chiral symmetry breaking (see \cite{Blask} \cite{Diakonov}) 

The model we have constructed,  however,  not only describes
the hadron mass spectrum; the basic Bethe-Salpeter amplitudes
we computed can be used to derive form factors and couplings of
various sorts \cite{Klaus}. Such a calculation has already been published for mesonic
states alone \cite{Koll}. In this paper we show now a more or less
complete calculation of electroweak nucleon form factors and
resonance transition form factors as far as experimental data are
available. We limit ourselves to non-strange baryons; baryons with
strangeness will be considered in another publication. 

The present paper is organized as follows: 

a) In the first section we
outline the theoretical derivation of our form factors. For more
details of the model itself we refer to our previous publications
\cite{Uli1}-\cite{Uli3}. Here we only describe the way, how current matrix elements
follow from the Bethe-Salpeter equation. Particular emphasis is put on
the treatment of the quark two-body interaction and its role in the
 three-body vertex function (see also appendix \ref{appendixA}).

b) In section 2 we show our results for the nucleonic electroweak form factors
in comparison to the experiment. Several theoretical aspects are
discussed in addition, in particular  the special effects of the quark
interaction and the relevance of  relativistic boosting. 

c) In section 3 we present our predictions for transition form
factors. We compare with various experimental data, which are,
however, not as well established as it is the case for the nucleon
properties, because the extraction of these form factors from
pion-photoproduction is highly model dependent. Less ambiguous are the
photon couplings (helicity amplitudes), for which we have indeed results, which show good
overall agreement with the known data.  The detailed behaviour of
these form factors as a function of $Q^2$ show, however, sometimes large
discrepancies with the functional behaviour extracted so far from
experiments. Our results can in this respect be regarded as alternative
predictions waiting for experimental verification.

At the end of this introduction we want to stress that the calculation
presented in this paper contains no free parameters or
normalization. All model parameters were fixed in the previous
calculation of the baryon mass spectrum \cite{Uli2} (We use the set of
model $\cal A$ from this reference which quantitatively describes
several features of the complete light flavor baryon spectrum.). We even did not try
to bring our form factor results into closer agreement to some
experimental values, when the disagreement was only due to slight
deviations of our model resonance masses from the known experimental
values, in order to be entirely consistent with our final goal to see
how far the constituent quark picture of hadron resonances works in
phenomenology at higher energies. 
	
\section{Current matrix elements derived from the Bethe-Salpeter amplitudes} 
\subsection{Bethe-Salpeter amplitudes}

In our first paper \cite{Uli1} we presented a formally covariant
constituent quark model of baryons which is based on the 
3-quark Bethe-Salpeter amplitudes 
\begin{eqnarray}
  \label{BS_amplitude}
  \chi_{\bar P\; a_1 a_2 a_3}(x_1,x_2,x_3)
  &:=&
  \VacBra\;
  T\;
  \Psi_{a_1}(x_1)
  \Psi_{a_2}(x_2)
  \Psi_{a_3}(x_3)\;
  \ket{\bar P}.\nn
\end{eqnarray}
In quantum field theory these are the transition matrix
elements of three quark field operators $\Psi_{a_i}(x_i)$ 
between a baryon state $\ket{\bar P}$ with 
mass $M= \sqrt{\bar P^2}$ and four momentum $\bar P = (\omega_{\bf P},{\bf P})= (\sqrt{|{\bf P}|^2 +M^2},{\bf P})$  
and the vacuum $\VacKet$.
The Fourier transform $\chi_{\bar P}(p_\xi,p_\eta)$
(here $p_\xi$ and and $p_\eta$ are the two relative Jacobi four-momenta)
formally obeys the Bethe-Salpeter equation
which in a short-hand operator notation
reads
\begin{equation}\label{BSEquation}
\chi_{\bar P} =  -\textrm{i}\;{G_0}_{\bar P}\;\left(K^{(3)}_{\bar P} + \overline K^{(2)}_{\bar P}\right)\;\chi_{\bar P}.
\end{equation}
Here ${G_0}_{P}$ denotes the three-fold tensor product 
\begin{eqnarray}
\label{G0_mom}
\lefteqn{{G_0}_{P}\;(p_\xi,p_\eta; p_\xi',p_\eta')=}\nn
&&S^1_F\!\left(\mbox{$\frac{1}{3}P\!+\!p_{\xi}\!+\!\frac{1}{2}p_{\eta}$}\right)
\ten
S^2_F\!\left(\mbox{$\frac{1}{3}P\!-\!p_{\xi}\!+\!\frac{1}{2}p_{\eta}$}\right)
\ten
S^3_F\!\left(\mbox{$\frac{1}{3}P\!-\!p_{\eta}$}\right)\nn
&& \hspace*{10mm}\times \;\;(2 \pi)^4\;\delta^{(4)}(p_\xi-p_\xi')\;\; (2 \pi)^4\;\delta^{(4)}(p_\eta-p_\eta')
\end{eqnarray}
of full quark-propagators $S^i_F$, $K^{(3)}_{P}$ is the irreducible three-body interaction kernel
and $\overline K^{(2)}_{P}$ the sum
\begin{eqnarray}
\label{K2_lift_fourier}
\lefteqn{\overline K^{(2)}_{P}(p_\xi,p_\eta;\;p_\xi',p_\eta')=}\nn
&&K^{(2)}_{(\frac{2}{3}P+p_{\eta})}(p_{\xi},p'_{\xi})\ten {S^3_F}^{-1}\left(\mbox{$\frac{1}{3}P-p_{\eta}$}\right)
(2\pi)^4 \; \delta^{(4)}(p_{\eta}-p'_{\eta})\nn
&&+ \quad\textrm{corresponding terms with interacting }\nn[-1mm] 
&&\phantom{+ \quad}\textrm{quark pairs (23) and (31)}
\end{eqnarray}
of the irreducible two-particle interactions
$K^{(2)}$ in each quark pair.\\

To be as close as possible in contact with the quite succesful
non-relativistic quark model, the  basic assumptions of this model 
are the following:
\begin{enumerate}
\item
The full quark propagators $S_F^i$ are replaced
by free Feynman propagators with effective constituent quark masses $m_i$:
\begin{equation}
\label{free_prop_approx}
S^i_{F}\left(p_i\right) 
\stackrel{!}{=} \frac{\rm i}{\not\! p_i - m_i + {\rm i}\epsilon }. 
\end{equation}
\item
We adopt instantaneous three- and two-body interaction kernels $K^{(3)}$
and $K^{(2)}$ which in the restframe of the baryon are described by unretarded
three- and two-body potentials $V^{(3)}$ and $V^{(2)}$:
\begin{eqnarray}
\label{inst_approx_CMS}
K^{(3)}_{P}(p_\xi,p_\eta;\;p_\xi',p_\eta')\bigg|_{P=(M,{\bf 0})}
&\stackrel{!}{=}&
V^{(3)}({\bf p_\xi},{\bf p_\eta};\;{\bf p_\xi'},{\bf p_\eta'}),\nn
\\
K^{(2)}_{(\frac{2}{3}P+p_{\eta})}(p_{\xi},p'_{\xi})
\bigg|_{ P=(M,{\bf 0})}
&\stackrel{!}{=}&
V^{(2)}({{\bf p}_{\xi}},{{\bf p}_{\xi}'}).\nonumber
\end{eqnarray}
In our specific quark model of baryons \cite{Uli2,Uli3} these potentials
represent string-like confinement for the three-particle kernel
and 't Hooft's instanton-induced interaction for the two-particle
kernel.
\end{enumerate}
These assumptions allow to eliminate the energy-like coordinates
$p_\xi^0$ and $p_\eta^0$ and thus to reduce the 3-Fermion Bethe-Salpeter equation to a
simpler equation -- known as Salpeter equation. 
In case of instantaneouse three-body forces alone this reduction
procedure is straightforward.
However, as discussed in
detail in ref. \cite{Uli1}, serious complications arise within
the reduction procedure, if genuine 2-body interactions are taken into
account in the three-body Bethe-Salpeter equation: the unconnected
two-body contribution $\overline K^{(2)}_P$ 
within the three-body system then prevents a straightforward reduction to
the Salpeter equation. In ref. \cite{Uli1} we presented a
method which -- in presence of a genuine instantaneous three-body kernel --
nevertheless allows a
reasonable treatment of these forces within the Salpeter framework.
There we derived a Salpeter equation for the (projected) Salpeter
amplitude (for a brief review see also appendix \ref{appendixA})
\begin{eqnarray}
\label{def:salpeter_ampl_CMS}
\Phi_M^\Lambda ({\bf p_\xi},{\bf p_\eta}) &:=&
\left(\Lambda^{+++}({\bf p_\xi},{\bf p_\eta}) + \Lambda^{---}({\bf p_\xi},{\bf p_\eta})\right)\nn
&&\times\;\int\frac{\textrm{d} p_\xi^0}{2\pi}\;\frac{\textrm{d} p_\eta^0}{2\pi}\;
\chi_{M}\left((p_\xi^0, {\bf p_\xi}), (p_\eta^0, {\bf p_\eta})\right),\nn
\end{eqnarray}
where
$\Lambda^{+++}({\bf p_\xi},{\bf p_\eta}):=\Lambda_1^+({\bf
p_1})\ten\Lambda_2^+({\bf p_2})\ten\Lambda_3^+({\bf p_3})$ and
$\Lambda^{---}({\bf p_\xi},{\bf p_\eta}):=\Lambda_1^-({\bf
p_1})\ten\Lambda_2^-({\bf p_2})\ten\Lambda_3^-({\bf p_3})$ are 
projection operators onto purely positive-energy and purely
negative-energy three-quark states, respectively.  This is achieved by
a perturbative elimination of retardation effects which arise due to
the two-body interaction. To this end we constructed an instantaneous
three-body kernel $V^{\rm eff}_{M}$ which effectively parameterizes
 the effects of the two-body forces.  We expanded this
quasi-potential in powers of a residual kernel $K^R_M= \overline
K^{(2)}_M + V^{(3)}_R$ which is the sum of the retarded two-body
contribution $\overline K^{(2)}_M$ and that part $V^{(3)}_R$ of the
instantaneous three-body kernel $V^{(3)}$ that couples to the
mixed energy components. In lowest order (Born approximation) $V^{\rm
eff}_{M}\approx {V^{\rm eff}_{M}}^{(1)}$ of this perturbative
expansion one finds \begin{eqnarray}
\lefteqn{
{V^{\rm eff}_{M}}^{(1)} ({\bf p_\xi}, {\bf p_\eta};\;{\bf p_\xi'}, {\bf p_\eta'})=}\nn[3mm]
&&(2 \pi)^3\;\delta^{(3)}({\bf p_\eta}\!-\!{\bf p_\eta'})\;\;\gamma^0\ten\gamma^0\ten\gamma^0\nn
&&
\Bigg[
\Lambda^{+++}({\bf p_\xi},{\bf p_\eta})\!\!
\left[\gamma^0\!\ten\!\gamma^0\;V^{(2)}({\bf p_\xi}, {\bf p_\xi'})\right]\!\ten\!\Id\;
\Lambda^{+++}({\bf p'_\xi},{\bf p_\eta})\nn
&&-
\Lambda^{---}({\bf p_\xi},{\bf p_\eta})\!\!
\left[\gamma^0\!\ten\!\gamma^0\;V^{(2)}({\bf p_\xi}, {\bf p_\xi'})\right]\!\ten\!\Id\;
\Lambda^{---}({\bf p'_\xi},{\bf p_\eta})\Bigg]\nn[3mm]
\label{Veff_Born_explicit}
&&
+ \quad\textrm{corresponding terms with interacting}\nn[-1mm] 
&&\phantom{+ \quad}\textrm{quark pairs (23) and (31),}
\end{eqnarray}
and the Salpeter equation then reads
explicitly as follows:
\begin{eqnarray}
\label{Salp_Hamilt_V3_V2_born}
\lefteqn{\Phi_M^\Lambda({\bf p_\xi}, {\bf p_\eta})=}\nn[3mm]
&& 
\left[
\frac{\Lambda^{+++}({\bf p_\xi},{\bf p_\eta})}{M\!-\!\Omega({\bf p_\xi}, {\bf p_\eta})\!+\!{\rm i}\epsilon}
+
\frac{\Lambda^{---}({\bf p_\xi},{\bf p_\eta})}{M\!+\!\Omega({\bf p_\xi}, {\bf p_\eta})\!-\!{\rm i}\epsilon}
\right]
\gamma^0\!\ten\!\gamma^0\!\ten\!\gamma^0\nn
&&\quad\times\;\int 
\frac{\textrm{d}^3 p_\xi'}{(2\pi)^3}\;
\frac{\textrm{d}^3 p_\eta'}{(2\pi)^3}\;
V^{(3)}({\bf p_\xi},{\bf p_\eta};\;{\bf p_\xi'},{\bf
  p_\eta'})\;
\Phi_M^\Lambda({\bf p_\xi'}, {\bf p_\eta'})\nn[3mm]
&+&
\left[
\frac{\Lambda^{+++}({\bf p_\xi},{\bf p_\eta})}{M\!-\!\Omega({\bf p_\xi}, {\bf p_\eta})\!+\!{\rm i}\epsilon}
-
\frac{\Lambda^{---}({\bf p_\xi},{\bf p_\eta})}{M\!+\!\Omega({\bf p_\xi}, {\bf p_\eta})\!-\!{\rm i}\epsilon}
\right]
\gamma^0\!\ten\!\gamma^0\!\ten\!\Id\nn
&&\quad\quad\times\;\int 
\frac{\textrm{d}^3 p_\xi'}{(2\pi)^3}\;
V^{(2)}({\bf p_\xi},{\bf p_\xi'})\ten\Id\;\;
\Phi_M^\Lambda({\bf p_\xi'}, {\bf p_\eta})\nn[3mm]
&&\qquad + \quad\textrm{corresponding terms with interacting}\nn[-1mm] 
&&\phantom{\qquad+ \quad}\textrm{quark pairs (23) and (31).}
\end{eqnarray}
Here $\Omega({\bf p_\xi}, {\bf p_\eta}):=\omega_1({\bf p_1})
+\omega_2({\bf p_2}) +\omega_3({\bf p_3})$ denotes the sum of the
kinetic energies $\omega_i({\bf p_i})=\sqrt{|{\bf p_i}|^2 + m_i^2}$ of each
quark. This equation determines the baryon masses $M$ and the
corresponding Salpeter amplitudes $\Phi_M^\Lambda$. The full
Bethe-Salpeter amplitude $\chi_{\bar P}$ in the corresponding order of approximation
is then determined from the Sal\-peter amplitude
$\Phi_M^\Lambda$ as follows (for details see appendix \ref{appendixA}),
\begin{equation}
\label{reconstructing_chi}
\chi_{\bar P} = 
\left[
{G_0}_{\bar P} 
- {\rm i}{G_0}_{\bar P}\left(
V_R^{(3)} 
+ 
\overline K^{(2)}_{\bar P}
- {V^{\rm eff}_{\bar P}}^{(1)}
\right){G_0}_{\bar P} 
\right]\Gamma_{\bar P}^\Lambda,
\end{equation}
where we introduced the vertex function $\Gamma_M^\Lambda$
according to
\begin{eqnarray}
\lefteqn{\Gamma_M^\Lambda({\bf p_\xi}, {\bf p_\eta})
:= 
-{\rm i}\;\int 
\frac{\textrm{d}^3 p_\xi'}{(2\pi)^3}\;
\frac{\textrm{d}^3 p_\eta'}{(2\pi)^3}\;}\nn
&&
\!\!\left[
V^{(3)}_{\Lambda}({\bf p_\xi},{\bf p_\eta};{\bf p_\xi'},{\bf p_\eta'})
\!+\!
{V^{\rm eff}_{M}}^{(1)}\!({\bf p_\xi},{\bf p_\eta};{\bf p_\xi'},{\bf p_\eta'})
\right]\!
\Phi_M^\Lambda({\bf p_\xi'}, {\bf p_\eta'}).\nn
\end{eqnarray}
Here the baryon four-momentum $\bar P = M = (M,\vec 0)$ is at
rest; for a general four-momentum $\bar P$ on the mass shell the
vertex function must be boosted by a suitable Lorentz
transformation in the obvious way.

\subsection{Current matrix elements}
The physically relevant bound-state matrix elements  of the current
operator $\bra{\bar P}j^\mu(x)\ket{\bar P'}$ with
$$
j^\mu(x)\;:=\; :\overline\Psi(x)\hat q \gamma^\mu\Psi(x):
$$
where $\hat q$ is the charge operator are calculated  as follows:
First consider the eight-point Green's function 
\begin{eqnarray}  \label{Def_7_Grennsf}
\lefteqn{G^\mu(x_1,x_2,x_3,x,x'_1,x'_2,x'_3) =}\nn
&&-\langle 0 | \: 
T \, \Psi(x_1) \Psi(x_2) \Psi(x_3) 
j^\mu(x)\; 
\overline\Psi(x'_1) \overline\Psi(x'_2) \overline\Psi(x'_3)\; 
| 0 \rangle.  \nonumber \\
\end{eqnarray}
Fixing the specific time ordering 
$$
\mbox{min} \{ x^0_1,x^0_2,x^0_3 \} > x^0 > \mbox{max} \{ x'^0_1,x'^0_2,x'^0_3 \}
$$
and inserting physical baryon states $\ket{\bar P}$ and $\ket{\bar
P'}$ in between, we find that the Fourier transform of this quantity
must have poles at $P^0=\omega_{\bf P}:=\sqrt{|{\bf P}|^2 +M^2}$ and
${P'}^0=\omega'_{\bf P'}:=\sqrt{|{\bf P'}|^2 +{M'}^2}$. In the
vincinity of these poles the Fourier transform $G_{P,P'}^\mu$ of $G^\mu$
has the form
\begin{eqnarray} 
\label{GPPprimeLaurent1}
\lefteqn{G_{P,P'}^\mu (p_\xi,p_\eta,p'_\xi,p'_\eta) =}\nn
&&  
\frac{1}{4\; \omega_{\bf P}\; \omega'_{\bf P'}}  
\frac{\chi_{\bar P}(p_\xi,p_\eta)}{(P^0-\omega_{\bf P}+{\rm i}\epsilon)  }\:  
\langle \bar P | {j^\mu}(0) | \bar P' \rangle  \:
\frac{\overline\chi_{\bar P'}(p'_\xi,p'_\eta) }{(P'^0-\omega'_{\bf P'}+{\rm
i}\epsilon)} \nonumber \\
&&\qquad + \quad \textrm{regular terms for}\; P^0\rightarrow \omega_{\bf P}\;\textrm{and}\;P'^0\rightarrow \omega'_{\bf P'}.\nn 
\end{eqnarray}
On the other hand minimal coupling
yields 
\begin{eqnarray} \label{GPPprime}
\lefteqn{ G^\mu_{P, P'} (p_\xi,p_\eta,p_\xi',p_\eta')=}\\[3mm]
&&
\int \frac{{\rm d}^4 p''_\xi}{(2\pi)^4}\:\frac{{\rm d}^4 p''_\eta}{(2\pi)^4}\:\frac{{\rm d}^4 p'''_\xi}{(2\pi)^4}\:\frac{{\rm d}^4 p'''_\eta}{(2\pi)^4} \; 
G_{P}(p_\xi,p_\eta; p''_\xi,p''_\eta)\nn[-1mm]
&&\hspace*{45mm}\times\; K_{P,P'}^\mu(p_\xi'',p_\eta'',p'''_\xi,p'''_\eta)\nn 
&&\hspace*{52mm}\times\; G_{P'}(p'''_\xi,p'''_\eta;p'_\xi,p'_\eta).\nonumber
\end{eqnarray}
Here $G_{P}$ is the  
Fourier transform of the six-point function 
\begin{eqnarray}
\label{greensfunction}
\lefteqn{G(x_1,x_2,x_3;x_1',x_2',x_3'):=}&&\\
& & 
-\VacExpect{\;
T\;
\Psi(x_1) \Psi(x_2) \Psi(x_3) 
\overline\Psi(x'_1) \overline\Psi(x'_2) \overline\Psi(x'_3)
}\nonumber.
\end{eqnarray}
$K_{P,P'}^\mu$ denotes the current kernel in momentum space which due
to the presence of a 2-body interaction kernel is given by the sum
\begin{equation}
\label{current_kernel}
K_{P,P'}^\mu= K_{P,P'}^{\mu (0)} + K_{P,P'}^{\mu (1)}  
\end{equation}
with 
\begin{eqnarray}
\label{current_kernel0}
\lefteqn{K_{P,P'}^{\mu (0)}(p_\xi,p_\eta,p'_\xi,p'_\eta)=}\nn
&&
{S^1_F}^{-1}\left(\mbox{$\frac{1}{3}P\!+\!p_{\xi}\!+\!\frac{1}{2}p_{\eta}$}\right)
\ten
{S^2_F}^{-1}\left(\mbox{$\frac{1}{3}P\!-\!p_{\xi}\!+\!\frac{1}{2}p_{\eta}$}\right)
\ten\;\hat q\; \gamma^\mu\nn
&& \times\; (2\pi)^4\delta^{(4)}\!(p_\xi-p'_\xi)\;
(2 \pi)^4\delta^{(4)}\!\!\left(\frac{2}{3}(P-P')+p_\eta-p_\eta'\right)\nn
&+& \textrm{corresponding photon couplings to quarks 1 and 2}\nn
\end{eqnarray}
and
\begin{eqnarray}
\label{current_kernel1}
\lefteqn{K_{P,P'}^{\mu (1)}(p_\xi,p_\eta,p'_\xi,p'_\eta)
=
{\rm i}\; K^{(2)}_{(\frac{2}{3}P+p_{\eta})}(p_{\xi},p'_{\xi})\;\ten\;\hat q\; \gamma^\mu}\nn
&& \qquad\qquad\qquad\quad \times\;(2 \pi)^4\;\delta^{(4)}\left(\frac{2}{3}(P-P')+p_\eta-p_\eta'\right)\nn
&+& \textrm{corresponding terms with photon couplings}\nn
& & \textrm{to quark 1 and quark 2.}
\end{eqnarray}
We now note (see ref. \cite{Uli1}) that the six-point Green's functions $G_{P}$
and $G_{P'}$ in the vincinity of the baryon energies 
$P^0=\omega_{\bf P}$ and ${P'}^0=\omega'_{\bf P'}$, respectively,
behave as
\begin{eqnarray}
G_P(p_\xi,p_\eta;p_\xi',p_\eta')
&=&
\frac{-\textrm{i}}{2\omega_{\bf P}}\;
\frac{\chi_{\bar P}(p_\xi,p_\eta)\;\overline\chi_{\bar P}(p'_\xi,p'_\eta)}
{P^0-\omega_{\bf P} + \textrm{i}\epsilon}\\
&& +\;\textrm{regular terms for } P^0\rightarrow \omega_{\bf P}\nonumber
\end{eqnarray}
and
\begin{eqnarray}
G_P'(p_\xi,p_\eta;p_\xi',p_\eta')
&=&
\frac{-\textrm{i}}{2\omega'_{\bf P'}}\;
\frac{\chi_{\bar P'}(p_\xi,p_\eta)\;\overline\chi_{\bar P'}(p'_\xi,p'_\eta)}
{P'^0-\omega'_{\bf P'} + \textrm{i}\epsilon}\\
&& +\;\textrm{regular terms for } P'^0\rightarrow \omega'_{\bf P'}\nonumber
\end{eqnarray}
Inserting these Laurent expansions of $G_{P}$
and $G_{P'}$ into the previous eq. (\ref{GPPprime})
yields

\begin{eqnarray} \label{GPPprimeLaurent2}
\lefteqn{ G^\mu_{P,P'} (p_\xi,p_\eta,p_\xi',p_\eta')  
= -\frac{1}{4 \;\omega_{\bf P}\; \omega'_{\bf P'}} 
\frac{\chi_{\bar P}(p_\xi,p_\eta)}{(P^0-\omega_{\bf P}+{\rm i}\epsilon) } }\nn[3mm] 
&&
\bigg[\int \frac{{\rm d}^4 p''_\xi}{(2\pi)^4}\:\frac{{\rm d}^4 p''_\eta}{(2\pi)^4}\:\frac{{\rm d}^4 p'''_\xi}{(2\pi)^4}\:\frac{{\rm d}^4 p'''_\eta}{(2\pi)^4}\nonumber \\ 
&& 
\qquad\overline\chi_{\bar P}(p''_\xi,p''_\eta) \: 
K_{\bar P,\bar P'}^\mu(p_\xi'',p_\eta'',p'''_\xi,p'''_\eta) \: 
\chi_{\bar P}(p'''_\xi,p'''_\eta)\bigg]\nn[3mm] 
&&\hspace*{50mm} \frac{\overline\chi_{\bar
P}(p'_\xi,p'_\eta)}{(P'^0-\omega'_{\bf P'}+{\rm i}\epsilon) }   \nonumber \\
&& \qquad + \quad \textrm{regular terms for}\; P^0\rightarrow \omega_{\bf P}\;\textrm{and}\;P'^0\rightarrow \omega'_{\bf P'}.\nn
\end{eqnarray}
The comparison with eq. (\ref{GPPprimeLaurent1}) shows that indeed the formula
\begin{eqnarray} \label{Matrix_Stromk}
\lefteqn{\langle \bar P | {j^\mu}(0) | \bar P' \rangle  = - \overline\chi_{\bar P}\;K_{\bar P, \bar P'}^\mu\;\chi_{\bar P'}=}\\  
&& - \int \frac{{\rm d}^4 p_\xi}{(2\pi)^4}\:\frac{{\rm d}^4 p_\eta}{(2\pi)^4}\:\frac{{\rm d}^4 p'_\xi}{(2\pi)^4}\:\frac{{\rm d}^4 p'_\eta}{(2\pi)^4}\nn 
&&\qquad\qquad\overline\chi_{\bar P}(p_\xi,p_\eta) \: K_{\bar P, \bar P'}^\mu(p_\xi,p_\eta,p'_\xi,p'_\eta) \: \chi_{\bar P'}(p'_\xi,p'_\eta)\nonumber
\end{eqnarray}
must hold. Our equations (\ref{reconstructing_chi}) and (\ref{current_kernel}) for $\chi_{\bar P}$ and
the current kernel $K_{\bar P, \bar P'}$ can now be inserted into this formula.
The result is 
\begin{equation}
\label{Matrix_Stromk_eff}
\langle \bar P | {j^\mu}(0) | \bar P' \rangle  
= 
- \overline\chi_{\bar P}\;K_{\bar P, \bar P'}^\mu\;\chi_{\bar P'} 
= 
- \overline\Gamma^\Lambda_{\bar P}\;\mathcal{K}_{\bar P, \bar P'}^\mu\;\Gamma^\Lambda_{\bar P'}
\end{equation}
where $\overline\Gamma^\Lambda_{\bar P}$ is the adjoint vertex
function which in the rest frame of the baryon is related to
$\Gamma^\Lambda_{\bar P}$ by 
\begin{equation}
{\overline\Gamma^\Lambda_{M}} =
-{\Gamma^\Lambda_{M}}^\dagger\gamma^0\ten\gamma^0\ten\gamma^0.
\end{equation}
$\mathcal{K}_{\bar P, \bar P'}^\mu$ defines the effective current
kernel
\begin{eqnarray}
\lefteqn{\mathcal{K}_{\bar P, \bar P'}^\mu :=}\\ 
&&
\left[
{G_0}_{\bar P} 
- {\rm i}\;{G_0}_{\bar P}\left(
V_R^{(3)} 
+ 
\overline K^{(2)}_{\bar P}
- {V^{\rm eff}_{\bar P}}^{(1)}
\right){G_0}_{\bar P} 
\right]\nn
&&\qquad\times\;K_{\bar P, \bar P'}^\mu\nn
&&\qquad\quad\times\left[
{G_0}_{\bar P'} 
- {\rm i}\;{G_0}_{\bar P'}\left(
V_R^{(3)} 
+ 
\overline K^{(2)}_{\bar P'}
- {V^{\rm eff}_{\bar P'}}^{(1)}
\right){G_0}_{\bar P'} 
\right]\nonumber
\end{eqnarray}
 which we expand to the same order in the residual kernel as the
effective kernel $V^{\rm eff}_{M}$ used in the Salpeter equation,
{\it i.e.} up to the  first order
\begin{equation}
\mathcal{K}_{\bar P, \bar P'}^\mu =
\mathcal{K}_{\bar P, \bar P'}^{\mu (0)}
+
\mathcal{K}_{\bar P, \bar P'}^{\mu (1)}
+
\textrm{higher orders.}
\end{equation} 
Here the 0th order contribution is defined as
\begin{equation}
\mathcal{K}_{\bar P, \bar P'}^{\mu (0)}
=
{G_0}_{\bar P}\;K_{\bar P, \bar P'}^{\mu (0)}\;{G_0}_{\bar P'}
\end{equation}
and reads explicitly
\begin{eqnarray}
\label{current_kernel_eff_0thOrder}
\lefteqn{\mathcal{K}_{\bar P,\bar P'}^{\mu (0)}(p_\xi,p_\eta,p'_\xi,p'_\eta)=}\nn[3mm]
&&
{S^1_F}\left(\mbox{$\frac{1}{3}\bar P\!+\!p_{\xi}\!+\!\frac{1}{2}p_{\eta}$}\right)
\ten
{S^2_F}\left(\mbox{$\frac{1}{3}\bar P\!-\!p_{\xi}\!+\!\frac{1}{2}p_{\eta}$}\right)\nn
&&
\qquad\qquad
\ten\;
{S^3_F}\left(\mbox{$\frac{1}{3}\bar P\!-\!p_{\eta}$}\right)
\hat q\; \gamma^\mu
{S^3_F}\left(\mbox{$\frac{1}{3}\bar P'\!-\!p'_{\eta}$}\right)
\nn
&& \times\; (2\pi)^4\delta^{(4)}(p_\xi-p'_\xi)\;
(2 \pi)^4\delta^{(4)}\!\left(\frac{2}{3}(\bar P-\bar P')+p_\eta-p_\eta'\right)\nn
&+& \textrm{corresponding photon couplings to quarks 1 and 2.}\nn
\end{eqnarray}
The 1st order contribution reads
\begin{eqnarray}
\lefteqn{
\mathcal{K}_{\bar P, \bar P'}^{\mu (1)}
=
{G_0}_{\bar P}\;K_{\bar P, \bar P'}^{\mu (1)}\;{G_0}_{\bar P'}}\nn
&&
- {\rm i}\;
{G_0}_{\bar P}\;
K_{\bar P, \bar P'}^{\mu (0)}\;
{G_0}_{\bar P'}\left(
V_R^{(3)} 
+ 
\overline K^{(2)}_{\bar P'}
- {V^{\rm eff}_{\bar P'}}^{(1)}
\right){G_0}_{\bar P'} \nn
&&- {\rm i}\;
{G_0}_{\bar P}\left(
V_R^{(3)} 
+ 
\overline K^{(2)}_{\bar P}
- {V^{\rm eff}_{\bar P}}^{(1)}
\right){G_0}_{\bar P}\;
K_{\bar P, \bar P'}^{\mu (0)}\;
{G_0}_{\bar P'}.\nn
\end{eqnarray}
Due to the complexity of the first order contribution
$\mathcal{K}_{\bar P, \bar P'}^{\mu (1)}$ to the current matrix
element we neglected so far this term in our calculations and took
only the 0th order term $\mathcal{K}_{\bar P, \bar P'}^{\mu (0)}$ into
account. We should mention at this point that in the static limit
$\bar P = \bar P' = M \equiv (M,{\bf 0})$ the first order term
does not contribute to the normalization of the charge as it vanishes
for the time-component of the current, {\it i.e.}
\begin{equation}
\overline\Gamma^\Lambda_{M}\;\mathcal{K}_{M, M}^{0 (1)}\;\Gamma^\Lambda_{M} = 0,
\end{equation}
while the 0th order term alone gives the correct normalization
due to the normalization condition of the Salpeter amplitudes.
Neglecting   $\mathcal{K}_{\bar P, \bar P'}^{\mu (1)}$
and setting the incoming baryon into the restframe, {\it i.e.}
$\bar P'= M \equiv (M, {\bf 0})$,
we then obtain for momentum transfer $q := \bar P - \bar P'$:
\begin{eqnarray}
\lefteqn{\langle \bar P | {j^\mu}(0) | M \rangle  
\simeq
- \overline\Gamma^\Lambda_{\bar P}\;\mathcal{K}_{\bar P, M}^{\mu (0)}\;\Gamma^\Lambda_{M}=}\nn
&& 
- 3\int \frac{{\rm d}^4 p_\xi}{(2\pi)^4}\:\frac{{\rm d}^4 p_\eta}{(2\pi)^4}\;
\overline\Gamma^\Lambda_{\bar P}(p_\xi,p_\eta-\frac{2}{3} q)\nn 
&&
{S^1_F}\left(\mbox{$\frac{1}{3}M\!+\!p_{\xi}\!+\!\frac{1}{2}p_{\eta}$}\right)
\ten
{S^2_F}\left(\mbox{$\frac{1}{3}M\!-\!p_{\xi}\!+\!\frac{1}{2}p_{\eta}$}\right)\nn
&&
\qquad
\ten\;
{S^3_F}\left(\mbox{$\frac{1}{3}M\!-\!p_{\eta}+q$}\right)
\hat q\; \gamma^\mu
{S^3_F}\left(\mbox{$\frac{1}{3}M\!-\!p_{\eta}$}\right)
\Gamma^\Lambda_{M}({\bf p_\xi},{\bf p_\eta}).\nonumber
\end{eqnarray}
This is the final expression for the current matrix elements which we use for the computation
of form factors in the next section.

\section{Electroweak form factors of the nucleon}

\begin{figure}[t]
\begingroup%
  \makeatletter%
  \newcommand{\GNUPLOTspecial}{%
    \@sanitize\catcode`\%=14\relax\special}%
  \setlength{\unitlength}{0.1bp}%
\begin{picture}(2339,2160)(0,0)%
\special{psfile=EFF_P llx=0 lly=0 urx=468 ury=504 rwi=4680}
\put(1876,1847){\makebox(0,0)[r]{\small Calc.}}%
\put(1876,1947){\makebox(0,0)[r]{\small MMD \cite{MMD}}}%
\put(1344,50){\makebox(0,0){$Q^2 [$GeV$^2]$}}%
\put(100,1180){%
\special{ps: gsave currentpoint currentpoint translate
270 rotate neg exch neg exch translate}%
\makebox(0,0)[b]{\shortstack{$G^p_E(Q^2)$}}%
\special{ps: currentpoint grestore moveto}%
}%
\put(2289,200){\makebox(0,0){3}}%
\put(1974,200){\makebox(0,0){2.5}}%
\put(1659,200){\makebox(0,0){2}}%
\put(1344,200){\makebox(0,0){1.5}}%
\put(1030,200){\makebox(0,0){1}}%
\put(715,200){\makebox(0,0){0.5}}%
\put(400,200){\makebox(0,0){0}}%
\put(350,2060){\makebox(0,0)[r]{1}}%
\put(350,1708){\makebox(0,0)[r]{0.8}}%
\put(350,1356){\makebox(0,0)[r]{0.6}}%
\put(350,1004){\makebox(0,0)[r]{0.4}}%
\put(350,652){\makebox(0,0)[r]{0.2}}%
\put(350,300){\makebox(0,0)[r]{0}}%
\end{picture}%
\endgroup
\caption{The electric form factor of the proton. Data are taken from
the compilation of P. Mergell et al. (MMD) \cite{MMD}. }\label{EFF_P}
\end{figure}
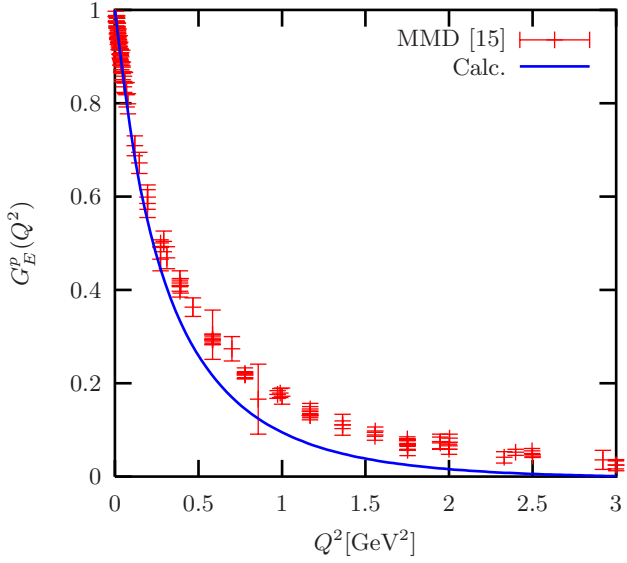

\begin{figure}[t]
\begingroup%
  \makeatletter%
  \newcommand{\GNUPLOTspecial}{%
    \@sanitize\catcode`\%=14\relax\special}%
  \setlength{\unitlength}{0.1bp}%
\begin{picture}(2339,2160)(0,0)%
\special{psfile=EFF_N llx=0 lly=0 urx=468 ury=504 rwi=4680}
\put(1876,1447){\makebox(0,0)[r]{\small Calc.}}%
\put(1876,1547){\makebox(0,0)[r]{\small Schiavilla\cite{Schiavilla}}}%
\put(1876,1647){\makebox(0,0)[r]{\small Eden\cite{Eden}}}%
\put(1876,1747){\makebox(0,0)[r]{\small Passchier \cite{Passchier}}}%
\put(1876,1847){\makebox(0,0)[r]{\small Herberg\cite{Herberg}, Ostrick\cite{Ostrick}}}%
\put(1876,1947){\makebox(0,0)[r]{\small Argonne V18}}%
\put(1369,50){\makebox(0,0){$Q^2 [$GeV$^2]$}}%
\put(100,1180){%
\special{ps: gsave currentpoint currentpoint translate
270 rotate neg exch neg exch translate}%
\makebox(0,0)[b]{\shortstack{$G^n_E(Q^2)$}}%
\special{ps: currentpoint grestore moveto}%
}%
\put(2289,200){\makebox(0,0){3}}%
\put(1983,200){\makebox(0,0){2.5}}%
\put(1676,200){\makebox(0,0){2}}%
\put(1370,200){\makebox(0,0){1.5}}%
\put(1063,200){\makebox(0,0){1}}%
\put(757,200){\makebox(0,0){0.5}}%
\put(450,200){\makebox(0,0){0}}%
\put(400,2060){\makebox(0,0)[r]{0.14}}%
\put(400,1809){\makebox(0,0)[r]{0.12}}%
\put(400,1557){\makebox(0,0)[r]{0.1}}%
\put(400,1306){\makebox(0,0)[r]{0.08}}%
\put(400,1054){\makebox(0,0)[r]{0.06}}%
\put(400,803){\makebox(0,0)[r]{0.04}}%
\put(400,551){\makebox(0,0)[r]{0.02}}%
\put(400,300){\makebox(0,0)[r]{0}}%
\end{picture}%
\endgroup
\caption{The electric form factor of the neutron.}\label{EFF_N}
\end{figure}
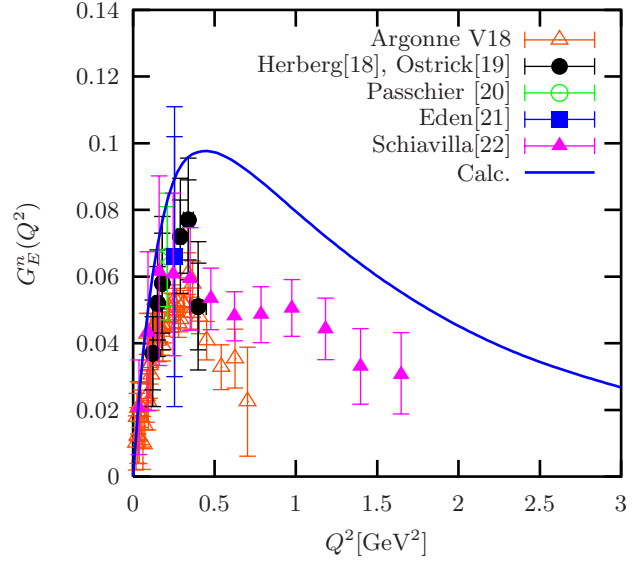

\begin{figure}[b]
\begingroup%
  \makeatletter%
  \newcommand{\GNUPLOTspecial}{%
    \@sanitize\catcode`\%=14\relax\special}%
  \setlength{\unitlength}{0.1bp}%
\begin{picture}(2339,2160)(0,0)%
\special{psfile=GE_over_GM llx=0 lly=0 urx=468 ury=504 rwi=4680}
\put(1876,1847){\makebox(0,0)[r]{\small Calc.}}%
\put(1876,1947){\makebox(0,0)[r]{\small Jones \cite{Jones}}}%
\put(1344,50){\makebox(0,0){$Q^2 [$GeV$^2]$}}%
\put(100,1180){%
\special{ps: gsave currentpoint currentpoint translate
270 rotate neg exch neg exch translate}%
\makebox(0,0)[b]{\shortstack{$\mu_pG^p_E/G^p_M(Q^2)$}}%
\special{ps: currentpoint grestore moveto}%
}%
\put(2289,200){\makebox(0,0){3}}%
\put(1974,200){\makebox(0,0){2.5}}%
\put(1659,200){\makebox(0,0){2}}%
\put(1344,200){\makebox(0,0){1.5}}%
\put(1030,200){\makebox(0,0){1}}%
\put(715,200){\makebox(0,0){0.5}}%
\put(400,200){\makebox(0,0){0}}%
\put(350,1900){\makebox(0,0)[r]{1}}%
\put(350,1580){\makebox(0,0)[r]{0.8}}%
\put(350,1260){\makebox(0,0)[r]{0.6}}%
\put(350,940){\makebox(0,0)[r]{0.4}}%
\put(350,620){\makebox(0,0)[r]{0.2}}%
\put(350,300){\makebox(0,0)[r]{0}}%
\end{picture}%
\endgroup
\caption{The ratio $\mu_pG^p_E/G^p_M$ compared with the new JLab data \cite{Jones}.}\label{EFF_MFF_P}
\end{figure}
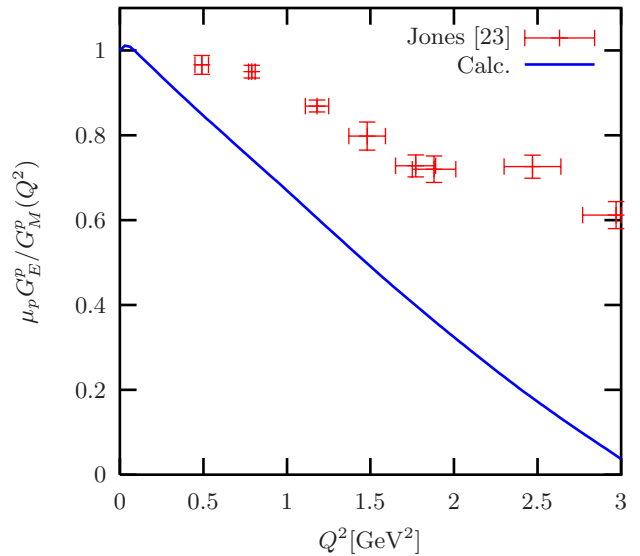

On the basis of the theoretical considerations of the preceeding
section we compute the current matrix elements
\beq
\langle{N,\bar P, \lambda}|j^{E,A}_\mu(0)|N,\bar P',\lambda'\rangle,
\eeq
for a nucleon state $|N,\bar P, \lambda\rangle$ with momentum $\bar P$ and helicity
$\lambda$, where $j^{E,A}_\mu(0)$ is the electromagnetic or  axial current
operator
\beq
j^E_\mu(x):= \overline\Psi(x)\hat q \gamma_\mu\Psi(x);\;  j^A_\mu(x):= \overline\Psi(x)\tau^+\!\!\gamma_\mu\!\gamma^5\Psi(x).
\eeq
They determine the electric, magnetic and
axial form factors according to ($Q^2 = - q^2$) 
\beqa
G^N_E(Q^2) := \frac{j^N_0(Q^2)}{\sqrt{4M^2+Q^2}}, & & G^N_M(Q^2) := \frac{j^N_+(Q^2)}{2\sqrt{Q^2}}
\eeqa
where
\beqa
j^N_0(0) & := & \langle{N,\bar P, \frac12}|j^E_0(0)|N,\bar P',\frac12\rangle\\
j^N_+(0) & := & \langle{N,\bar P, \frac12}|j^E_1(0)|N,\bar P',-\frac12\rangle\nonumber\\
	 &    & + i \langle{N,\bar P, \frac12}|j^E_2(0)|N,\bar P',-\frac12\rangle
\eeqa
for the electromagnetic form factors and
\begin{figure}[t]
\begingroup%
  \makeatletter%
  \newcommand{\GNUPLOTspecial}{%
    \@sanitize\catcode`\%=14\relax\special}%
  \setlength{\unitlength}{0.1bp}%
\begin{picture}(2339,2160)(0,0)%
\special{psfile=EFF_nT_P_N llx=0 lly=0 urx=468 ury=504 rwi=4680}
\put(1344,50){\makebox(0,0){$Q^2 [$GeV$^2]$}}%
\put(100,1180){%
\special{ps: gsave currentpoint currentpoint translate
270 rotate neg exch neg exch translate}%
\makebox(0,0)[b]{\shortstack{$G^{p/n}_E(Q^2)$}}%
\special{ps: currentpoint grestore moveto}%
}%
\put(2289,200){\makebox(0,0){3}}%
\put(1974,200){\makebox(0,0){2.5}}%
\put(1659,200){\makebox(0,0){2}}%
\put(1344,200){\makebox(0,0){1.5}}%
\put(1030,200){\makebox(0,0){1}}%
\put(715,200){\makebox(0,0){0.5}}%
\put(400,200){\makebox(0,0){0}}%
\put(350,2060){\makebox(0,0)[r]{1}}%
\put(350,1740){\makebox(0,0)[r]{0.8}}%
\put(350,1420){\makebox(0,0)[r]{0.6}}%
\put(350,1100){\makebox(0,0)[r]{0.4}}%
\put(350,780){\makebox(0,0)[r]{0.2}}%
\put(350,460){\makebox(0,0)[r]{0}}%
\end{picture}%
\endgroup
\caption{The electric form factor of the proton (solid) and neutron (dashed) calculated without the
instanton induced residual interaction compared with the experimental dipole shape $G_D$(dashed-dotted).}\label{EFF_nT}
\end{figure}
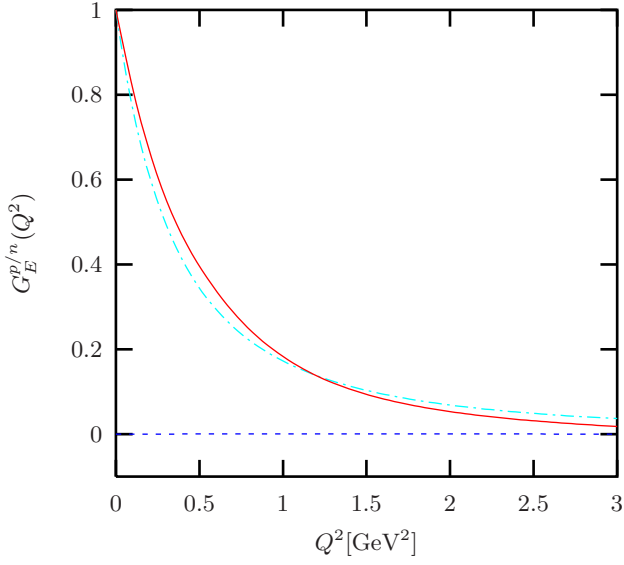
\begin{figure}[b]
\begingroup%
  \makeatletter%
  \newcommand{\GNUPLOTspecial}{%
    \@sanitize\catcode`\%=14\relax\special}%
  \setlength{\unitlength}{0.1bp}%
\begin{picture}(2339,2160)(0,0)%
\special{psfile=MFF_P llx=0 lly=0 urx=468 ury=504 rwi=4680}
\put(1876,1847){\makebox(0,0)[r]{\small Calc.}}%
\put(1876,1947){\makebox(0,0)[r]{\small MMD \cite{MMD}}}%
\put(1344,50){\makebox(0,0){$Q^2 [$GeV$^2]$}}%
\put(100,1180){%
\special{ps: gsave currentpoint currentpoint translate
270 rotate neg exch neg exch translate}%
\makebox(0,0)[b]{\shortstack{$G^p_M(Q^2)$}}%
\special{ps: currentpoint grestore moveto}%
}%
\put(2289,200){\makebox(0,0){3}}%
\put(1974,200){\makebox(0,0){2.5}}%
\put(1659,200){\makebox(0,0){2}}%
\put(1344,200){\makebox(0,0){1.5}}%
\put(1030,200){\makebox(0,0){1}}%
\put(715,200){\makebox(0,0){0.5}}%
\put(400,200){\makebox(0,0){0}}%
\put(350,1950){\makebox(0,0)[r]{3}}%
\put(350,1675){\makebox(0,0)[r]{2.5}}%
\put(350,1400){\makebox(0,0)[r]{2}}%
\put(350,1125){\makebox(0,0)[r]{1.5}}%
\put(350,850){\makebox(0,0)[r]{1}}%
\put(350,575){\makebox(0,0)[r]{0.5}}%
\put(350,300){\makebox(0,0)[r]{0}}%
\end{picture}%
\endgroup
\caption{The magnetic form factor of the proton. Data are taken from
the compilation of P. Mergell et al. (MMD) \cite{MMD}.}\label{MFF_P}
\end{figure}
\begin{figure}[t]
\begingroup%
  \makeatletter%
  \newcommand{\GNUPLOTspecial}{%
    \@sanitize\catcode`\%=14\relax\special}%
  \setlength{\unitlength}{0.1bp}%
\begin{picture}(2339,2160)(0,0)%
\special{psfile=EFF_isoscalvec llx=0 lly=0 urx=468 ury=504 rwi=4680}
\put(1344,50){\makebox(0,0){$Q^2 [$GeV$^2]$}}%
\put(100,1180){%
\special{ps: gsave currentpoint currentpoint translate
270 rotate neg exch neg exch translate}%
\makebox(0,0)[b]{\shortstack{$G^{s/v}_E(Q^2)$}}%
\special{ps: currentpoint grestore moveto}%
}%
\put(2289,200){\makebox(0,0){3}}%
\put(1974,200){\makebox(0,0){2.5}}%
\put(1659,200){\makebox(0,0){2}}%
\put(1344,200){\makebox(0,0){1.5}}%
\put(1030,200){\makebox(0,0){1}}%
\put(715,200){\makebox(0,0){0.5}}%
\put(400,200){\makebox(0,0){0}}%
\put(350,2060){\makebox(0,0)[r]{1}}%
\put(350,1740){\makebox(0,0)[r]{0.8}}%
\put(350,1420){\makebox(0,0)[r]{0.6}}%
\put(350,1100){\makebox(0,0)[r]{0.4}}%
\put(350,780){\makebox(0,0)[r]{0.2}}%
\put(350,460){\makebox(0,0)[r]{0}}%
\end{picture}%
\endgroup
\caption{The electric isoscalar (solid) and isovector (dashed) form
factors fo the nucleon compared with the experimental dipole shape $G_D$(dashed-dotted).}\label{isoscalvec}
\end{figure}
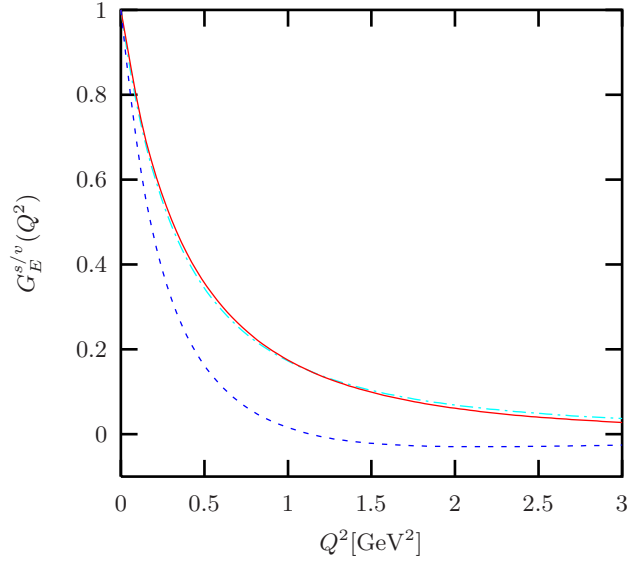
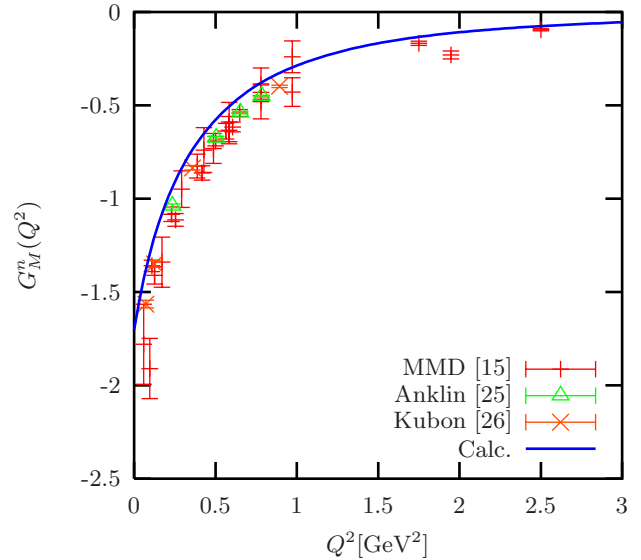
\begin{figure}[b]
\begingroup%
  \makeatletter%
  \newcommand{\GNUPLOTspecial}{%
    \@sanitize\catcode`\%=14\relax\special}%
  \setlength{\unitlength}{0.1bp}%
\begin{picture}(2339,2160)(0,0)%
\special{psfile=MFF_N llx=0 lly=0 urx=468 ury=504 rwi=4680}
\put(1876,413){\makebox(0,0)[r]{\small Calc.}}%
\put(1876,513){\makebox(0,0)[r]{\small Kubon \cite{Kubon}}}%
\put(1876,613){\makebox(0,0)[r]{\small Anklin \cite{Anklin}}}%
\put(1876,713){\makebox(0,0)[r]{\small MMD \cite{MMD}}}%
\put(1369,50){\makebox(0,0){$Q^2 [$GeV$^2]$}}%
\put(100,1180){%
\special{ps: gsave currentpoint currentpoint translate
270 rotate neg exch neg exch translate}%
\makebox(0,0)[b]{\shortstack{$G^n_M(Q^2)$}}%
\special{ps: currentpoint grestore moveto}%
}%
\put(2289,200){\makebox(0,0){3}}%
\put(1983,200){\makebox(0,0){2.5}}%
\put(1676,200){\makebox(0,0){2}}%
\put(1370,200){\makebox(0,0){1.5}}%
\put(1063,200){\makebox(0,0){1}}%
\put(757,200){\makebox(0,0){0.5}}%
\put(450,200){\makebox(0,0){0}}%
\put(400,2060){\makebox(0,0)[r]{0}}%
\put(400,1708){\makebox(0,0)[r]{-0.5}}%
\put(400,1356){\makebox(0,0)[r]{-1}}%
\put(400,1004){\makebox(0,0)[r]{-1.5}}%
\put(400,652){\makebox(0,0)[r]{-2}}%
\put(400,300){\makebox(0,0)[r]{-2.5}}%
\end{picture}%
\endgroup
\caption{The magnetic  form factor of the neutron compared with (MMD)\cite{MMD} and new data from MAMI\cite{Anklin}\cite{Kubon}.}\label{MFF_N}
\end{figure}
\beqa
G_A(Q^2) :=  \frac{j_A^+(Q^2)}{\sqrt{4M^2+Q^2}}
\eeqa
with 
\beqa
j_A^+(Q^2) & := & \langle{p,\bar P, \frac12}|j^A_1(0)|n,\bar P',-\frac12\rangle\nonumber\\
	   &    & + i \langle{p,\bar P, \frac12}|j^A_2(0)|n,\bar P',-\frac12\rangle
\eeqa
for the axial one. The normalization of the form factors is such that
the static magnetic moments and the axial coupling are given by
\beqa
\mu_M := G_M(Q^2=0), & & g_A:= G_A(Q^2=0).
\eeqa

The result for the proton electric form factor is shown in
fig. \ref{EFF_P}. The form factor obviously falls off too
rapidly. The electric form factor of the neutron, shown in
fig. \ref{EFF_N}, rises sharply; the high $Q^2$ behaviour of our
theoretical prediction is still acceptable. The sharp rise is in
qualitative agreement with more recent data, but our result overshoots
the experimental values. The rapid fall off of the proton form factor
and the sharp rise of the neutron form factor result from the action of
the 't Hooft interaction. To demonstrate this we show in
fig.  \ref{EFF_nT} the result of a calculation with the confinement
force alone (which, of course, will not yield a satisfactory
spectrum).

A more closer look at the results shows that it is the behaviour of
the isovector form factor $G^v_E := G^p_E - G^n_E$, which is
responsible for the disagreement with the empirical data. The
isoscalar form factor $G^s_E := G^p_E + G^n_E$ shows indeed even a
perfect dipole behaviour (see fig. \ref{isoscalvec}), consistent with
the experimental parametrization $G_D(Q^2) =
(1+Q^2/0.71\mbox{GeV}^2)^{-2}$. We want to note in addition that the
neutron form 
factor, which we have computed, still has a chance to agree with
experiment, because the extraction from deuteron scattering is not
free of ambiguities. A recent paper \cite{Tomasi}, which treats this
problem, produces in fact neutron form factors in qualitative
agreement with our results.

\begin{figure}[t]
\begingroup%
  \makeatletter%
  \newcommand{\GNUPLOTspecial}{%
    \@sanitize\catcode`\%=14\relax\special}%
  \setlength{\unitlength}{0.1bp}%
\begin{picture}(2339,2160)(0,0)%
\special{psfile=GA_ref llx=0 lly=0 urx=468 ury=504 rwi=4680}
\put(1876,1447){\makebox(0,0)[r]{Calc.}}%
\put(1876,1547){\makebox(0,0)[r]{\small Bloom \cite{Bloom}}}%
\put(1876,1647){\makebox(0,0)[r]{\small Joos \cite{Joos}}}%
\put(1876,1747){\makebox(0,0)[r]{\small Amaldi \cite{Amaldi}}}%
\put(1876,1847){\makebox(0,0)[r]{\small Brauel \cite{Brauel}}}%
\put(1876,1947){\makebox(0,0)[r]{\small delGuerra \cite{delGuerra}}}%
\put(1344,50){\makebox(0,0){$Q^2 [$GeV$^2]$}}%
\put(100,1180){%
\special{ps: gsave currentpoint currentpoint translate
270 rotate neg exch neg exch translate}%
\makebox(0,0)[b]{\shortstack{$G_A(Q^2)$}}%
\special{ps: currentpoint grestore moveto}%
}%
\put(2117,200){\makebox(0,0){1}}%
\put(1774,200){\makebox(0,0){0.8}}%
\put(1430,200){\makebox(0,0){0.6}}%
\put(1087,200){\makebox(0,0){0.4}}%
\put(743,200){\makebox(0,0){0.2}}%
\put(400,200){\makebox(0,0){0}}%
\put(350,1943){\makebox(0,0)[r]{1.4}}%
\put(350,1708){\makebox(0,0)[r]{1.2}}%
\put(350,1473){\makebox(0,0)[r]{1}}%
\put(350,1239){\makebox(0,0)[r]{0.8}}%
\put(350,1004){\makebox(0,0)[r]{0.6}}%
\put(350,769){\makebox(0,0)[r]{0.4}}%
\put(350,535){\makebox(0,0)[r]{0.2}}%
\put(350,300){\makebox(0,0)[r]{0}}%
\end{picture}%
\endgroup
\caption{The axial   form factor of the nucleon. Data are taken from
the compilation of V. Bernard et al. \cite{bernard}.}\label{axial}
\end{figure}
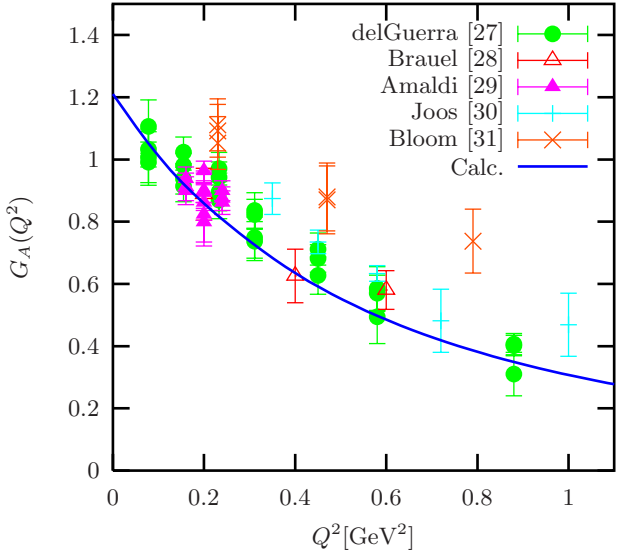

\begin{table}[b]
\begin{center}
\caption{Static properties of the nucleon.\label{static}}
\begin{tabular}{cccc}
\hline\noalign{\smallskip}
					& Calc.		& Exp. \\
\noalign{\smallskip}\hline\noalign{\smallskip}
$\mu_p$					& $2.74$ $ \mu_N$	& $2.793$ $\mu_N$  	& \cite{PDG2k}\\
$\mu_n$					& $-1.70$ $\mu_N$	& $-1.913$ $\mu_N$ 	& \cite{PDG2k}\\
$\sqrt{\langle r^2\rangle^p_E}$		& $0.82$ fm	& $0.847$ fm     		& \cite{MMD}\\
${\langle r^2\rangle^n_E}$		& $0.11$ fm$^2$	& $0.113\pm 0.004$ fm$^2$ 	& \cite{MMD}\\
$\sqrt{\langle r^2\rangle^p_M}$		& $0.91$ fm	& $0.836$ fm     		& \cite{MMD}\\
$\sqrt{\langle r^2\rangle^n_M}$		& $0.86$ fm	& $0.889$ fm     		& \cite{MMD}\\
$g_A$					& $1.21$	& $1.2670 \pm 0.0035$		& \cite{PDG2k}\\
$\sqrt{\langle r^2\rangle_A}$		& $0.62$ fm	& $0.61 \pm 0.01$ fm     	& \cite{bernard}\\
\noalign{\smallskip}\hline
\end{tabular}
\end{center}
\end{table}
In figs. \ref{MFF_P} and \ref{MFF_N} we show our results for the magnetic form
factors of proton and neutron. Obviously we describe the data very
well. This is interesting in so far as we induce by 't Hooft's force
strong correlations in the amplitudes, which in standard
non-relativistic quark models spoil the symmetry of the wave function
and destroy therefore the classical SU(6) -results for magnetic
moments. We believe that the correct relativistic boosting of our
amplitudes is responsible for the good agreement with the data. For
comparison we have calculated the magnetic moment of the proton
omitting the boost (as in non-relativistic calculations). The value
drops in fact by the order of one magneton.

Very recently the ratio $\mu_pG^p_E/G^p_M$ has been measured with
high accuracy at Jefferson Lab \cite{Jones}. The data show a
monotonical, almost linear  decrease with increasing $Q^2$ indicating that the
electric form factor of the proton decreases significantly faster than
the dipole $G_D$, which is in qualitative agreement with our results
(see fig. \ref{EFF_MFF_P}). But due to the rapid fall off of $G^p_E$
the ratio is strongly underestimated in our model for high $Q^2$.

Figure \ref{axial} shows our result for the axial form factor in
comparison with the experimental data, which show in fact large
deviations between several experimental groups. Our theoretical
results agree, however, very well with the more recent experimental
work.

We conclude this section with a table of static electroweak constants
of the nucleon (table \ref{static}). Apparently we achieved a reasonably good agreement
with the common experimental values.

\section{Transition form factors}
\begin{figure}[t]
\begingroup%
  \makeatletter%
  \newcommand{\GNUPLOTspecial}{%
    \@sanitize\catcode`\%=14\relax\special}%
  \setlength{\unitlength}{0.1bp}%
\begin{picture}(2339,2160)(0,0)%
\special{psfile=P33_Ng llx=0 lly=0 urx=468 ury=504 rwi=4680}
\put(1876,1547){\makebox(0,0)[r]{\small  Calc.}}%
\put(1876,1647){\makebox(0,0)[r]{\small Frolov \cite{Frolov}}}%
\put(1876,1747){\makebox(0,0)[r]{\small  Stein \cite{Stein}}}%
\put(1876,1847){\makebox(0,0)[r]{\small  Foster \cite{Foster}}}%
\put(1876,1947){\makebox(0,0)[r]{\small  Bartel \cite{Bartel}}}%
\put(1344,50){\makebox(0,0){$Q^2 [$GeV$^2]$}}%
\put(100,1180){%
\special{ps: gsave currentpoint currentpoint translate
270 rotate neg exch neg exch translate}%
\makebox(0,0)[b]{\shortstack{$G_M^*(Q^2)$}}%
\special{ps: currentpoint grestore moveto}%
}%
\put(2244,200){\makebox(0,0){4}}%
\put(2019,200){\makebox(0,0){3.5}}%
\put(1794,200){\makebox(0,0){3}}%
\put(1569,200){\makebox(0,0){2.5}}%
\put(1345,200){\makebox(0,0){2}}%
\put(1120,200){\makebox(0,0){1.5}}%
\put(895,200){\makebox(0,0){1}}%
\put(670,200){\makebox(0,0){0.5}}%
\put(445,200){\makebox(0,0){0}}%
\put(350,2003){\makebox(0,0)[r]{3}}%
\put(350,1719){\makebox(0,0)[r]{2.5}}%
\put(350,1435){\makebox(0,0)[r]{2}}%
\put(350,1152){\makebox(0,0)[r]{1.5}}%
\put(350,868){\makebox(0,0)[r]{1}}%
\put(350,584){\makebox(0,0)[r]{0.5}}%
\put(350,300){\makebox(0,0)[r]{0}}%
\end{picture}%
\endgroup
\caption{$\Delta(1232)$ magnetic transition form factor.}\label{Del_N}
\end{figure}
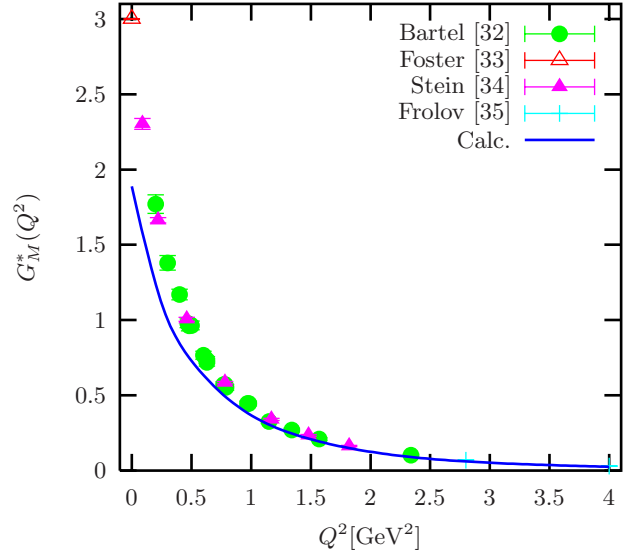


The nucleon-$\Delta$ transition form factor is intensively studied since many
years. A long standing problem is the smallness of this quantity at
low $Q^2$ in  quark model calculations, which can only be cured by hybrid
models with a mesonic cloud around the nucleon \cite{Sato}\cite{Kamalov}. Our result,
compared to the experimental data, is shown  in fig. \ref{Del_N}. We see that at
small $Q^2$ we do not cure this old quark model prediction and that
our form factor remains too small. The $Q^2$ behaviour above 1 GeV$^2$, however, is
correct.

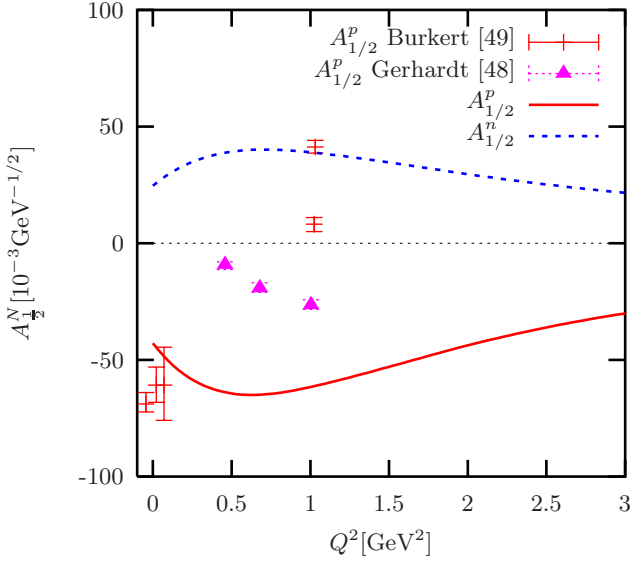
\begin{figure}[t]
\begingroup%
  \makeatletter%
  \newcommand{\GNUPLOTspecial}{%
    \@sanitize\catcode`\%=14\relax\special}%
  \setlength{\unitlength}{0.1bp}%
\begin{picture}(2339,2160)(0,0)%
\special{psfile=P11_Ng llx=0 lly=0 urx=468 ury=504 rwi=4680}
\put(1876,1584){\makebox(0,0)[r]{\small $A^n_{1/2}$}}%
\put(1876,1702){\makebox(0,0)[r]{\small $A^p_{1/2}$}}%
\put(1876,1820){\makebox(0,0)[r]{\small $A^p_{1/2}$ Gerhardt \cite{Gerhardt}}}%
\put(1876,1938){\makebox(0,0)[r]{\small $A^p_{1/2}$ Burkert \cite{Burkert}}}%
\put(1369,50){\makebox(0,0){$Q^2 [$GeV$^2]$}}%
\put(100,1180){%
\special{ps: gsave currentpoint currentpoint translate
270 rotate neg exch neg exch translate}%
\makebox(0,0)[b]{\shortstack{$A^N_{\frac 12} [10^{-3}$GeV$^{-1/2}]$}}%
\special{ps: currentpoint grestore moveto}%
}%
\put(2289,200){\makebox(0,0){3}}%
\put(1992,200){\makebox(0,0){2.5}}%
\put(1696,200){\makebox(0,0){2}}%
\put(1399,200){\makebox(0,0){1.5}}%
\put(1103,200){\makebox(0,0){1}}%
\put(806,200){\makebox(0,0){0.5}}%
\put(509,200){\makebox(0,0){0}}%
\put(400,2060){\makebox(0,0)[r]{100}}%
\put(400,1620){\makebox(0,0)[r]{50}}%
\put(400,1180){\makebox(0,0)[r]{0}}%
\put(400,740){\makebox(0,0)[r]{-50}}%
\put(400,300){\makebox(0,0)[r]{-100}}%
\end{picture}%
\endgroup
\caption{$P_{11}(1440)$ electroexcitation helicity amplitudes.}\label{roper}
\end{figure}

\begin{figure}[b]
\begingroup%
  \makeatletter%
  \newcommand{\GNUPLOTspecial}{%
    \@sanitize\catcode`\%=14\relax\special}%
  \setlength{\unitlength}{0.1bp}%
\begin{picture}(2339,2160)(0,0)%
\special{psfile=D13_Ng llx=0 lly=0 urx=468 ury=504 rwi=4680}
\put(1876,1348){\makebox(0,0)[r]{$A^n_{3/2}$}}%
\put(1876,1466){\makebox(0,0)[r]{$A^n_{1/2}$}}%
\put(1876,1584){\makebox(0,0)[r]{$A^p_{3/2}$}}%
\put(1876,1702){\makebox(0,0)[r]{$A^p_{1/2}$}}%
\put(1876,1820){\makebox(0,0)[r]{\small $A^p_{3/2}$  Burkert \cite{Burkert}}}%
\put(1876,1938){\makebox(0,0)[r]{\small $A^p_{1/2}$ Burkert \cite{Burkert}}}%
\put(1369,50){\makebox(0,0){$Q^2 [$GeV$^2]$}}%
\put(100,1180){%
\special{ps: gsave currentpoint currentpoint translate
270 rotate neg exch neg exch translate}%
\makebox(0,0)[b]{\shortstack{$A^N_{\lambda} [10^{-3}$GeV$^{-1/2}]$}}%
\special{ps: currentpoint grestore moveto}%
}%
\put(2232,200){\makebox(0,0){3}}%
\put(1944,200){\makebox(0,0){2.5}}%
\put(1657,200){\makebox(0,0){2}}%
\put(1370,200){\makebox(0,0){1.5}}%
\put(1082,200){\makebox(0,0){1}}%
\put(795,200){\makebox(0,0){0.5}}%
\put(507,200){\makebox(0,0){0}}%
\put(400,2060){\makebox(0,0)[r]{200}}%
\put(400,1840){\makebox(0,0)[r]{150}}%
\put(400,1620){\makebox(0,0)[r]{100}}%
\put(400,1400){\makebox(0,0)[r]{50}}%
\put(400,1180){\makebox(0,0)[r]{0}}%
\put(400,960){\makebox(0,0)[r]{-50}}%
\put(400,740){\makebox(0,0)[r]{-100}}%
\put(400,520){\makebox(0,0)[r]{-150}}%
\put(400,300){\makebox(0,0)[r]{-200}}%
\end{picture}%
\endgroup
\caption{$D_{13}(1520)$ electroexcitation helicity amplitudes.}\label{D13}
\end{figure}

\begin{figure}[t]
\begingroup%
  \makeatletter%
  \newcommand{\GNUPLOTspecial}{%
    \@sanitize\catcode`\%=14\relax\special}%
  \setlength{\unitlength}{0.1bp}%
\begin{picture}(2339,2160)(0,0)%
\special{psfile=S11_Ng llx=0 lly=0 urx=468 ury=504 rwi=4680}
\put(1876,1247){\makebox(0,0)[r]{\small $A^n_{1/2}$}}%
\put(1876,1347){\makebox(0,0)[r]{\small $A^p_{1/2}$}}%
\put(1876,1447){\makebox(0,0)[r]{\small  Capstick \cite{Capstick2}}}%
\put(1876,1547){\makebox(0,0)[r]{\small \cite{Breuker}$-$\cite{Kummer}}}%
\put(1876,1647){\makebox(0,0)[r]{\small  Brasse \cite{Brasse84}}}%
\put(1876,1747){\makebox(0,0)[r]{\small  Krusche \cite{Krusche}}}%
\put(1876,1847){\makebox(0,0)[r]{\small  Armstrong \cite{Jlab}}}%
\put(1876,1947){\makebox(0,0)[r]{\small  Thompson \cite{CLAS}}}%
\put(1369,50){\makebox(0,0){$Q^2 [$GeV$^2]$}}%
\put(100,1180){%
\special{ps: gsave currentpoint currentpoint translate
270 rotate neg exch neg exch translate}%
\makebox(0,0)[b]{\shortstack{$A^N_{\frac 12} [10^{-3}$GeV$^{-1/2}]$}}%
\special{ps: currentpoint grestore moveto}%
}%
\put(2289,200){\makebox(0,0){4}}%
\put(2065,200){\makebox(0,0){3.5}}%
\put(1840,200){\makebox(0,0){3}}%
\put(1616,200){\makebox(0,0){2.5}}%
\put(1392,200){\makebox(0,0){2}}%
\put(1168,200){\makebox(0,0){1.5}}%
\put(943,200){\makebox(0,0){1}}%
\put(719,200){\makebox(0,0){0.5}}%
\put(495,200){\makebox(0,0){0}}%
\put(400,2060){\makebox(0,0)[r]{250}}%
\put(400,1809){\makebox(0,0)[r]{200}}%
\put(400,1557){\makebox(0,0)[r]{150}}%
\put(400,1306){\makebox(0,0)[r]{100}}%
\put(400,1054){\makebox(0,0)[r]{50}}%
\put(400,803){\makebox(0,0)[r]{0}}%
\put(400,551){\makebox(0,0)[r]{-50}}%
\put(400,300){\makebox(0,0)[r]{-100}}%
\end{picture}%
\endgroup
\caption{$S_{11}(1535)$ electroexcitation helicity amplitudes. }\label{S11_1535}
\end{figure}

\begin{figure}[b]
\begingroup%
  \makeatletter%
  \newcommand{\GNUPLOTspecial}{%
    \@sanitize\catcode`\%=14\relax\special}%
  \setlength{\unitlength}{0.1bp}%
\begin{picture}(2339,2160)(0,0)%
\special{psfile=S11_1650_Ng llx=0 lly=0 urx=468 ury=504 rwi=4680}
\put(1876,1747){\makebox(0,0)[r]{$A^n_{1/2}$}}%
\put(1876,1847){\makebox(0,0)[r]{$A^p_{1/2}$}}%
\put(1876,1947){\makebox(0,0)[r]{\small $A^p_{1/2}$ Burkert \cite{Burkert}}}%
\put(1344,50){\makebox(0,0){$Q^2 [$GeV$^2]$}}%
\put(100,1180){%
\special{ps: gsave currentpoint currentpoint translate
270 rotate neg exch neg exch translate}%
\makebox(0,0)[b]{\shortstack{$A^N_{\frac 12} [10^{-3}$GeV$^{-1/2}]$}}%
\special{ps: currentpoint grestore moveto}%
}%
\put(2289,200){\makebox(0,0){3}}%
\put(1984,200){\makebox(0,0){2.5}}%
\put(1680,200){\makebox(0,0){2}}%
\put(1375,200){\makebox(0,0){1.5}}%
\put(1070,200){\makebox(0,0){1}}%
\put(766,200){\makebox(0,0){0.5}}%
\put(461,200){\makebox(0,0){0}}%
\put(350,2060){\makebox(0,0)[r]{80}}%
\put(350,1809){\makebox(0,0)[r]{60}}%
\put(350,1557){\makebox(0,0)[r]{40}}%
\put(350,1306){\makebox(0,0)[r]{20}}%
\put(350,1054){\makebox(0,0)[r]{0}}%
\put(350,803){\makebox(0,0)[r]{-20}}%
\put(350,551){\makebox(0,0)[r]{-40}}%
\put(350,300){\makebox(0,0)[r]{-60}}%
\end{picture}%
\endgroup
\caption{$S_{11}(1650)$ electroexcitation helicity amplitudes. }\label{S11_1650}
\end{figure}

\begin{figure}[t]
\begingroup%
  \makeatletter%
  \newcommand{\GNUPLOTspecial}{%
    \@sanitize\catcode`\%=14\relax\special}%
  \setlength{\unitlength}{0.1bp}%
\begin{picture}(2339,2160)(0,0)%
\special{psfile=D13_1700_Ng llx=0 lly=0 urx=468 ury=504 rwi=4680}
\put(1876,1584){\makebox(0,0)[r]{$A^n_{3/2}$}}%
\put(1876,1702){\makebox(0,0)[r]{$A^n_{1/2}$}}%
\put(1876,1820){\makebox(0,0)[r]{$A^p_{3/2}$}}%
\put(1876,1938){\makebox(0,0)[r]{$A^p_{1/2}$}}%
\put(1344,50){\makebox(0,0){$Q^2 [$GeV$^2]$}}%
\put(100,1180){%
\special{ps: gsave currentpoint currentpoint translate
270 rotate neg exch neg exch translate}%
\makebox(0,0)[b]{\shortstack{$A^N_{\lambda} [10^{-3}$GeV$^{-1/2}]$}}%
\special{ps: currentpoint grestore moveto}%
}%
\put(2230,200){\makebox(0,0){3}}%
\put(1935,200){\makebox(0,0){2.5}}%
\put(1640,200){\makebox(0,0){2}}%
\put(1345,200){\makebox(0,0){1.5}}%
\put(1049,200){\makebox(0,0){1}}%
\put(754,200){\makebox(0,0){0.5}}%
\put(459,200){\makebox(0,0){0}}%
\put(350,2060){\makebox(0,0)[r]{80}}%
\put(350,1809){\makebox(0,0)[r]{60}}%
\put(350,1557){\makebox(0,0)[r]{40}}%
\put(350,1306){\makebox(0,0)[r]{20}}%
\put(350,1054){\makebox(0,0)[r]{0}}%
\put(350,803){\makebox(0,0)[r]{-20}}%
\put(350,551){\makebox(0,0)[r]{-40}}%
\put(350,300){\makebox(0,0)[r]{-60}}%
\end{picture}%
\endgroup
\caption{$D_{13}(1700)$ electroexcitation helicity amplitudes.}\label{D13_1700}
\end{figure}

\begin{figure}[b]
\begingroup%
  \makeatletter%
  \newcommand{\GNUPLOTspecial}{%
    \@sanitize\catcode`\%=14\relax\special}%
  \setlength{\unitlength}{0.1bp}%
\begin{picture}(2339,2160)(0,0)%
\special{psfile=S31_Ng llx=0 lly=0 urx=468 ury=504 rwi=4680}
\put(1876,1847){\makebox(0,0)[r]{\small $A^N_{1/2}$}}%
\put(1876,1947){\makebox(0,0)[r]{\small $A^N_{1/2}$ Burkert \cite{Burkert}}}%
\put(1344,50){\makebox(0,0){$Q^2 [$GeV$^2]$}}%
\put(100,1180){%
\special{ps: gsave currentpoint currentpoint translate
270 rotate neg exch neg exch translate}%
\makebox(0,0)[b]{\shortstack{$A^N_{\frac 12} [10^{-3}$GeV$^{-1/2}]$}}%
\special{ps: currentpoint grestore moveto}%
}%
\put(2289,200){\makebox(0,0){3}}%
\put(1984,200){\makebox(0,0){2.5}}%
\put(1680,200){\makebox(0,0){2}}%
\put(1375,200){\makebox(0,0){1.5}}%
\put(1070,200){\makebox(0,0){1}}%
\put(766,200){\makebox(0,0){0.5}}%
\put(461,200){\makebox(0,0){0}}%
\put(350,1884){\makebox(0,0)[r]{40}}%
\put(350,1532){\makebox(0,0)[r]{20}}%
\put(350,1180){\makebox(0,0)[r]{0}}%
\put(350,828){\makebox(0,0)[r]{-20}}%
\put(350,476){\makebox(0,0)[r]{-40}}%
\end{picture}%
\endgroup
\caption{$S_{31}(1620)$ electroexcitation helicity amplitude.}\label{S31_1620}
\end{figure}
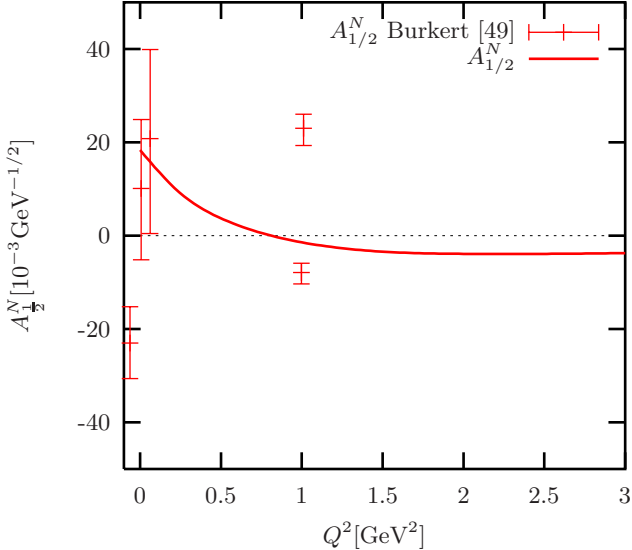

\begin{figure}[t]
\begingroup%
  \makeatletter%
  \newcommand{\GNUPLOTspecial}{%
    \@sanitize\catcode`\%=14\relax\special}%
  \setlength{\unitlength}{0.1bp}%
\begin{picture}(2339,2160)(0,0)%
\special{psfile=D15_Ng llx=0 lly=0 urx=468 ury=504 rwi=4680}
\put(1876,1584){\makebox(0,0)[r]{$A^n_{3/2}$}}%
\put(1876,1702){\makebox(0,0)[r]{$A^n_{1/2}$}}%
\put(1876,1820){\makebox(0,0)[r]{$A^p_{3/2}$}}%
\put(1876,1938){\makebox(0,0)[r]{$A^p_{1/2}$}}%
\put(1344,50){\makebox(0,0){$Q^2 [$GeV$^2]$}}%
\put(100,1180){%
\special{ps: gsave currentpoint currentpoint translate
270 rotate neg exch neg exch translate}%
\makebox(0,0)[b]{\shortstack{$A^N_{\lambda} [10^{-3}$GeV$^{-1/2}]$}}%
\special{ps: currentpoint grestore moveto}%
}%
\put(2230,200){\makebox(0,0){3}}%
\put(1935,200){\makebox(0,0){2.5}}%
\put(1640,200){\makebox(0,0){2}}%
\put(1345,200){\makebox(0,0){1.5}}%
\put(1049,200){\makebox(0,0){1}}%
\put(754,200){\makebox(0,0){0.5}}%
\put(459,200){\makebox(0,0){0}}%
\put(350,2060){\makebox(0,0)[r]{80}}%
\put(350,1809){\makebox(0,0)[r]{60}}%
\put(350,1557){\makebox(0,0)[r]{40}}%
\put(350,1306){\makebox(0,0)[r]{20}}%
\put(350,1054){\makebox(0,0)[r]{0}}%
\put(350,803){\makebox(0,0)[r]{-20}}%
\put(350,551){\makebox(0,0)[r]{-40}}%
\put(350,300){\makebox(0,0)[r]{-60}}%
\end{picture}%
\endgroup
\caption{$D_{15}(1675)$ electroexcitation helicity amplitudes.}\label{D15_1675}
\end{figure}
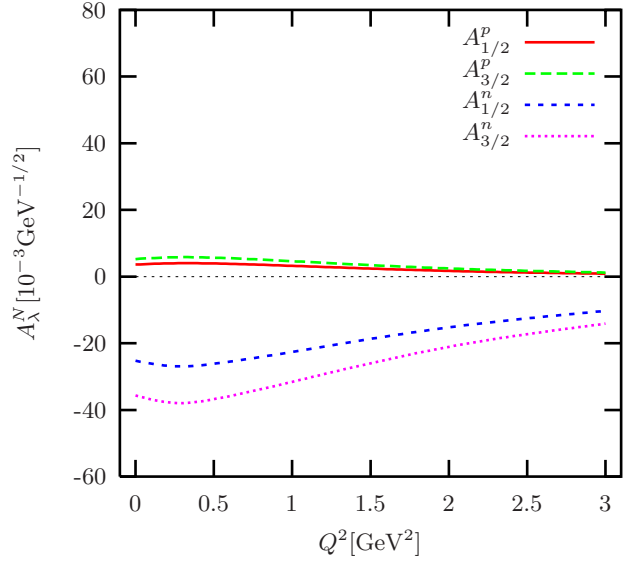

\begin{figure}[b]
\begingroup%
  \makeatletter%
  \newcommand{\GNUPLOTspecial}{%
    \@sanitize\catcode`\%=14\relax\special}%
  \setlength{\unitlength}{0.1bp}%
\begin{picture}(2339,2160)(0,0)%
\special{psfile=D33_Ng llx=0 lly=0 urx=468 ury=504 rwi=4680}
\put(1876,1584){\makebox(0,0)[r]{$A^N_{3/2}$}}%
\put(1876,1702){\makebox(0,0)[r]{$A^N_{1/2}$}}%
\put(1876,1820){\makebox(0,0)[r]{\small $A^N_{3/2}$  Burkert \cite{Burkert}}}%
\put(1876,1938){\makebox(0,0)[r]{\small $A^N_{1/2}$ Burkert \cite{Burkert}}}%
\put(1344,50){\makebox(0,0){$Q^2 [$GeV$^2]$}}%
\put(100,1180){%
\special{ps: gsave currentpoint currentpoint translate
270 rotate neg exch neg exch translate}%
\makebox(0,0)[b]{\shortstack{$A^N_{\lambda} [10^{-3}$GeV$^{-1/2}]$}}%
\special{ps: currentpoint grestore moveto}%
}%
\put(2230,200){\makebox(0,0){3}}%
\put(1935,200){\makebox(0,0){2.5}}%
\put(1640,200){\makebox(0,0){2}}%
\put(1345,200){\makebox(0,0){1.5}}%
\put(1049,200){\makebox(0,0){1}}%
\put(754,200){\makebox(0,0){0.5}}%
\put(459,200){\makebox(0,0){0}}%
\put(350,2060){\makebox(0,0)[r]{200}}%
\put(350,1620){\makebox(0,0)[r]{150}}%
\put(350,1180){\makebox(0,0)[r]{100}}%
\put(350,740){\makebox(0,0)[r]{50}}%
\put(350,300){\makebox(0,0)[r]{0}}%
\end{picture}%
\endgroup
\caption{$D_{33}(1700)$ electroexcitation helicity amplitudes.}\label{D33_1700}
\end{figure}

The results for the transition form factors of the second and third resonance
region are presented in the form of helicity amplitudes $A^p_\lambda$ and
$A^n_\lambda$ for proton and neutron targets, respectively, defined by
\beqa
A^N_\lambda & := & C\langle N^*,\bar P,\lambda|j^E_1(0)+ij^E_2(0)|N,\bar P',\lambda - 1\rangle
\eeqa
with $C := \sqrt{\frac{\pi\alpha}{2M_{N^*}(M_{N^*}^2-M_N^2)}}$
following \cite{Capstick1}. For the Roper resonance $P_{11}(1440)$ (see
fig. \ref{roper}) experimental data
 are unfortunately contradicting each other, because the
extraction from pion-photoproduction is strongly model dependent. The
experimental situation is much better for the $S_{11}(1535)$ and
$D_{13}(1520)$, though not completely conclusive. In fig. \ref{S11_1535} we
show the available experimental data for the $S_{11}$ together with
 our result and the result of a  quark model calculation by
 Capstick \cite{Capstick2}. We see 
that our model works better at low $Q^2$ but  our form factor possibly
drops too fast. The helicity amplitude $A^p_{1/2}$ for
$D_{13}(1520)$ (see fig. \ref{D13}) is well reproduced, but the amplitude $A^p_{3/2}$
seems to be too small even at the photon point. To complete our
results for the $1\hbar\omega$ shell we show the helicity
amplitudes for $S_{11}(1650)$, $D_{13}(1700)$ and
$D_{15}(1675)$ in figs. \ref{S11_1650},\ref{D13_1700} and \ref{D15_1675} and for the
$\Delta$ resonances $S_{31}(1620)$ and $D_{33}(1700)$ in
figs. \ref{S31_1620} and \ref{D33_1700}. Experimental
 data to compare with, taken from  \cite{Burkert}, are again quite
contradictory but new measurements at Jefferson Laboratory are in progress.

The photon couplings of all these resonances are summarized in table
\ref{Photonc} and, with the exceptions discussed just before, agree
quite well with the data. Our results for the second and third resonance region
are therefore quite satis\-factory; because of the total lack of
experimental data we stop, however, our investigation of form factors at
this point, though we could in principle continue with higher resonances.

\begin{table}[t]
\begin{center}
\caption{Photon couplings.\label{Photonc}}
\begin{tabular}{lc@{\hspace{0.2em}}r@{\hspace{0.6em}}rc@{\hspace{0.2em}}r@{\hspace{0.6em}}r}
\hline\noalign{\smallskip}
\mbox{state}		&			& \mbox{Calc.}		&\mbox{Exp. \cite{PDG2k}}	&                       & \mbox{Calc.}            &\mbox{Exp. \cite{PDG2k}}     \\
\noalign{\smallskip}\hline\noalign{\smallskip}

$P_{33} (1232)$		& $A^N_{\frac12}$	& $-89	$		& $-135 \pm 6$	\\
	      		& $A^N_{\frac32}$	& $-152	$		& $-255 \pm 8$	\\
\noalign{\smallskip}\hline\noalign{\smallskip}
$S_{11}(1535) $		& $A^p_{\frac12}$	& $113	$		& $90 \pm 30$	& $A^n_{\frac12}$       &  $-75	$       & $-46\pm 27$   \\
$S_{11}(1650) $		& $A^p_{\frac12}$	& $5	$		& $53 \pm 16$	& $A^n_{\frac12}$       &  $-16$        & $-15\pm 21$   \\
$D_{13}(1520) $		& $A^p_{\frac12}$	& $-53	$		& $-24 \pm 9$	& $A^n_{\frac12}$       &  $1$          & $-59\pm 9$   \\
	      		& $A^p_{\frac32}$	& $51	$		& $166 \pm 5$	& $A^n_{\frac32}$       &  $-52$         & $-139\pm 11$   \\
$D_{13}(1700) $		& $A^p_{\frac12}$	& $-13	$		& $-18 \pm 13$	& $A^n_{\frac12}$       &  $16$	        & $0   \pm 50$  \\
	      		& $A^p_{\frac32}$	& $-10	$		& $-2 \pm 24$	& $A^n_{\frac32}$       &  $-42$        & $-3 \pm 44$   \\
$D_{15}(1675) $		& $A^p_{\frac12}$	& $4	$		& $19 \pm 8$	& $A^n_{\frac12}$       &  $-25$        & $-43\pm 12$    \\
	      		& $A^p_{\frac32}$	& $5	$		& $15 \pm 9$	& $A^n_{\frac32}$       &  $-33$        & $-58\pm 13$    \\
\noalign{\smallskip}\hline\noalign{\smallskip}
$P_{11}(1440)$		& $A^p_{\frac12}$ 	& $-48$			& $-65 \pm 4$	& $A^n_{\frac12}$       &  $27$         & $40  \pm 10$   \\
\noalign{\smallskip}\hline
$S_{31}(1620) $		& $A^N_{\frac12}$	& $18	$		& $27 \pm 11$	\\
$D_{33}(1700) $		& $A^N_{\frac12}$	& $63	$		& $104\pm 15$   \\
	      		& $A^N_{\frac32}$	& $68	$		& $ 85 \pm 22$	 \\
\noalign{\smallskip}\hline
\end{tabular}
\end{center}
\end{table}

\section{Conclusion}
On the basis of the Bethe-Salpeter equation we have computed nucleon
form factors and photon transition form factors of baryons up to the
third resonance region. Our results are in quantitative agreement with
the existent experimental data, but need further experimental
verification. Our fully relativistic treatment proofed to be
absolutely necessary to reach these results; in addition we were able
to demonstrate that our dynamical assumptions about the effective
quark forces at least do not lead to contradictions. In future work
higher orders of the effective kernels $V^{\rm eff}_{M}$ and
$\mathcal{K}_{\bar P, \bar P'}^\mu$ neglected so far will be taken
into account. So far we did not
find strong indications that the concept of constituent quarks fails
completely at the energies considered. There is of course room for
improvements, {\it e.g.} sea quark admixtures or pion cloud effects as used
in some hybrid models. We have made no efforts in this direction,
because our goal is to explore the concept of constituent quarks at
higher energies in order to find out, when it  really
fails. In the same spirit we are now performing similar calculations of
strange baryon properies and strong two-body decays of baryon
resonances. 
\begin{acknowledgement}
We want to thank our colleagues U. Mei\ss ner, E.Klempt,
F.Klein, B.Schoch, W. Pfeil, V.V. Anisovich, A. Sarantsev, H. Schmieden and
S. Capstick for many helpfull discussions and usefull hints. The
financial aid of the Deutsche Forschungsgemeinschaft is gratefully
acknowledged. 
\end{acknowledgement}
\appendix{\section{Appendix: Reconstruction of the Bethe-Salpeter amplitude}
\label{appendixA}
In our first paper \cite{Uli1} we demonstrated how, under
assumptions (\ref{free_prop_approx}) and (\ref{inst_approx_CMS}), the
Bethe-Salpeter equation (\ref{BSEquation}) 
\begin{equation}
\label{BSEquation2}
\chi_{M} =  -\textrm{i}\;{G_0}_{M}\;\left(V^{(3)} + \overline K^{(2)}_{M}\right)\;\chi_{M},
\end{equation}
can be reduced (in the rest frame of the
baryon) to a Salpeter equation for the (projected)
Salpeter amplitude
\begin{eqnarray}
\label{def:salpeter_ampl_CMS_2}
\lefteqn{\Phi_M^\Lambda ({\bf p_\xi},{\bf p_\eta}) :=}\nn
&&\Lambda({\bf p_\xi},{\bf p_\eta})\;\int\frac{\textrm{d} p_\xi^0}{2\pi}\;\frac{\textrm{d} p_\eta^0}{2\pi}\;
\chi_{M}\left((p_\xi^0, {\bf p_\xi}), (p_\eta^0, {\bf p_\eta})\right).
\end{eqnarray}
Here
\begin{eqnarray}
\Lambda({\bf p_\xi},{\bf p_\eta})&:=&
\Lambda^{+++}({\bf p_\xi},{\bf p_\eta}) + \Lambda^{---}({\bf p_\xi},{\bf p_\eta})\nn
&=&\phantom{+\;}
\Lambda_1^+({\bf p_1})\ten\Lambda_2^+({\bf p_2})\ten\Lambda_3^+({\bf p_3})\nn
&&+\;
\Lambda_1^-({\bf p_1})\ten\Lambda_2^-({\bf p_2})\ten\Lambda_3^-({\bf p_3})
\end{eqnarray}
is a projector on purely positive-energy and negative-energy Dirac
spinors.  To perform the reduction we split the integral kernel $K_M= V^{(3)} +
\overline K_M^{(2)}$ of the Bethe-Salpeter equation into two parts
\begin{equation}
K_M = V^{(3)}_\Lambda + (V^{(3)}_R + \overline K_M^{(2)}).
\end{equation}
The first part $V^{(3)}_\Lambda:= \overline\Lambda
\;V^{(3)}\Lambda$ of the kernel 
(with $\overline \Lambda := \gamma^0\ten\gamma^0\ten\gamma^0\;\Lambda\;\gamma^0\ten\gamma^0\ten\gamma^0$) 
is the particular contribution of the instantaneous three-body
potential $V^{(3)}$ which couples to purely positive-energy and
negative-energy components only. The second, residual part
$V^{(3)}_R + \overline K_M^{(2)}$ is the sum of the retarded two body
contribution $\overline K_M^{(2)}$ and the remaining  part $V^{(3)}_R := V^{(3)}-V^{(3)}_\Lambda$ of $V^{(3)}$ that
also couples to the mixed energy components. Putting the difficult residual contribution
into the resolvent 
\begin{equation}
\label{inhom_IE_GP}
{\cal G}_{M}
=
{G_0}_{M} -\textrm{i}\;{G_0}_{M}\; \left[V^{(3)}_{\rm R} + \overline K^{(2)}_{M}\right]\;{\cal G}_{M},
\end{equation}
the Bethe-Salpeter equation can be written in the form  
\begin{eqnarray}
\label{reconstruct_SA_Lam}
\chi_{M} 
&=&  
-\textrm{i}\;{\cal G}_{M}\;  V^{(3)}_\Lambda\;\Phi_{M}^{\Lambda}.
\end{eqnarray}
This form, firstly, gives a prescription how to reconstruct the full
Bethe-Salpeter amplitude $\chi_{M}$ from the Salpeter amplitude
$\Phi_{M}^{\Lambda}$ and, secondly, is suitable for the reduction to
the Salpeter equation as $V^{(3)}_\Lambda$ is instantaneous. We obtained
\begin{eqnarray}
\label{SE_V2_V3_cov}
\Phi_{M}^\Lambda &=&  
-\textrm{i}\;\langle{G_0}_{M}\rangle\;\left[V^{(3)}_\Lambda + V^{\rm
 eff}_{M}\right]\;\Phi^\Lambda_{M}
\end{eqnarray}
where the brackets $\langle\;\rangle$ denote the six-dimensional reduction
\begin{eqnarray}
\label{reduc_sixpoint}
\lefteqn {\langle A\rangle ({\bf p_\xi},{\bf p_\eta};\;{\bf p_\xi'},{\bf p_\eta'}):=}\nn 
&&\quad\quad\int\frac{\textrm{d} p_\xi^0}{2\pi}\;
\frac{\textrm{d} p_\eta^0}{2\pi}\;
\int\frac{\textrm{d} p_\xi'^0}{2\pi}\;
\frac{\textrm{d} p_\eta'^0}{2\pi}\;
A (p_\xi,p_\eta;\;p_\xi',p_\eta')
\end{eqnarray}
of any eight-dimensional six-point function $A$ and $V^{\rm eff}_{M}$
(with the property $\overline\Lambda\;V^{\rm eff}_{M}=V^{\rm eff}_{M}\;\Lambda=V^{\rm eff}_{M}$)
is an additional instantaneous three-body kernel which effectively
parameterizes the effects of the retarded two-body forces. The latter
is defined as the irreducible kernel for the resolvent $\langle{\cal
G}_{M}\rangle_\Lambda :=
\Lambda\langle{\cal G}_{M}\rangle\overline \Lambda$, where irreducible
is understood with respect to $\langle{G_0}_{M}\rangle$, {\it i.e.}
\begin{equation}
\label{reduced_IE_Veff}
\langle{\cal G}_{M}\rangle_\Lambda
\;\stackrel{!}{=}\;
\langle{G_0}_{M}\rangle 
-\textrm{i}\;\langle{G_0}_{M}\rangle\;V^{\rm eff}_{M}\;
\langle{\cal G}_{M}\rangle_\Lambda.
\end{equation}
To determine this quasi-potential we expanded it in powers
$k$ of the residual kernel $V^{(3)}_R + \overline K_M^{(2)}$, {\it i.e.}
\begin{equation}
\label{series_ansatz_Veff}
V^{\rm eff}_{M}
\;=\;
\sum_{k=1}^\infty  {V^{\rm eff}_{M}}^{(k)}.
\end{equation}
In ref. \cite{Uli1} we derived a generic formula to calculate the
terms ${V^{\rm eff}_{M}}^{(k)}$ of the series to arbitrary orders.  In
practice, however, we have to approximate the effective kernel $V^{\rm
eff}_{M}$, which consists of an infinite number of terms. A
systematical approximation is now given by truncating the series
(\ref{series_ansatz_Veff}) at some finite order $k < \infty$, {\it i.e.}
\begin{equation}
\label{Veff_series_approx}
V^{\rm eff}_{M} 
\simeq 
{V^{\rm eff}_{M}}^{(1)}
+
{V^{\rm eff}_{M}}^{(2)}
+
\ldots
+
{V^{\rm eff}_{M}}^{(k)},
\end{equation}
thus yielding an approximation of the  Salpeter
amplitude $\Phi^\Lambda_M \simeq{\Phi^\Lambda_M}^{\!\!(k)}$ by the solution of
\begin{equation}
\label{SEquation_approx}
{\Phi^\Lambda_M}^{\!\!(k)}
=-{\rm i}\langle {G_0}_M\rangle\left(V^{(3)}_\Lambda+\sum_{i=1}^k {V^{\rm eff}_{M}}^{(i)} \right){\Phi^\Lambda_M}^{\!\!(k)}.
\end{equation}
For the calculation of transition matrix elements we need the full
Bethe-Salpeter amplitude  which (if $V^{\rm eff}_{M}$ and
$\Phi^\Lambda_M$ are known exactly) can be reconstructed by the
prescription (\ref{reconstruct_SA_Lam}) via the Green's function
${\cal G}_M$.  However, the truncation of the Salpeter equation has the
consequence that relation (\ref{reconstruct_SA_Lam}) does not hold for
${\Phi^\Lambda_M}^{\!\!(k)}$.  To be consistent we need an
approximation of this reconstruction formula that corresponds to the
approximation (\ref{Veff_series_approx}) of the effective kernel. In
other words, we require the corresponding $k$th order  approximation
$\chi_{M}^{(k)}$ of the Bethe-Salpeter amplitude $\chi_{M}$ to be such that
its reduction according to eq. (\ref{def:salpeter_ampl_CMS_2}) yields
the $k$th order  approximation ${\Phi^\Lambda_M}^{\!\!(k)}$ of the
Salpeter amplitude. Here we want to show that a consistent
prescription for an approximated reconstruction of the Bethe-Salpeter
amplitude can indeed be found.  To this end, we
recast the Bethe-Salpeter and Salpeter equation into a more convenient
form.  We start with the exact Bethe-Salpeter equation
(\ref{BSEquation2}) and isolate the instantaneous part
$V^{(3)}_\Lambda+\sum_{i=1}^k {V^{\rm eff}_{M}}^{(i)}$ of the kernel
which enters in the $k$th order approximation of the Salpeter equation
(\ref{SEquation_approx}):
\begin{eqnarray}
\label{BSEquation3}
\chi_{M} &=&  -\textrm{i}\;{G_0}_{M}\;
\Bigg[\quad\;
\left(V_\Lambda^{(3)} + \sum_{i=1}^k {V^{\rm eff}_{M}}^{(i)}\right)\\[-1mm]
&&\qquad\qquad\; +\;\; \left(\overline K^{(2)}_{M} +  V_R^{(3)} - \sum_{i=1}^k {V^{\rm eff}_{M}}^{(i)}\right)
\Bigg]\;\chi_{M}.\nonumber
\end{eqnarray}
The exact Bethe-Salpeter equation can then be rewritten as follows
\begin{eqnarray}
\label{BSEquation4}
\chi_{M} &=&
-\textrm{i}\;{\cal G}^{R,k}_{M}\;\left(V_\Lambda^{(3)} + \sum_{i=1}^k {V^{\rm eff}_{M}}^{(i)}\right)\;\chi_{M}\\
\label{BSEquation5}
&&-\textrm{i}\;{\cal G}^{R,k}_{M}\;\left(V_\Lambda^{(3)} + \sum_{i=1}^k {V^{\rm eff}_{M}}^{(i)}\right)\;\Phi^\Lambda_{M},
\end{eqnarray}
giving a reconstruction formula for $\chi_{M}$ from $\Phi^\Lambda_{M}$,
which is equivalent to eq. (\ref{reconstruct_SA_Lam}) but better
suited to formulate our approximation for the full amplitude. Here a
new residual propagator ${\cal G}^{R,k}_{M}$ appears which describes
the propagation of the three quarks via the second part of the kernel
in (\ref{BSEquation3}) and which is defined by the integral equation
\begin{equation}
\label{GRk}
{\cal G}^{R,k}_{M} 
= {G_0}_{M} 
- 
\textrm{i}\;{G_0}_{M}\;\left(\overline K^{(2)}_{M} +  V_R^{(3)} - \sum_{i=1}^k {V^{\rm eff}_{M}}^{(i)}\right)\;{\cal G}^{R,k}_{M}.
\end{equation}
Performing the reduction of eq. (\ref{BSEquation5}) --  {\it i.e.} integration over the $p_\xi^0$,$p_\eta^0$-coordinates and multiplication 
with the projector $\Lambda$ --  we obtain the (still exact) Salpeter
equation in the form
\begin{eqnarray}
\label{SE_new}
\Phi_{M}^\Lambda &=&  
-\textrm{i}\;\langle{\cal G}^{R,k}_{M}\rangle_\Lambda\;\left[V^{(3)}_\Lambda 
+ \sum_{i=1}^k {V^{\rm eff}_{M}}^{(i)}\right]\;\Phi^\Lambda_{M}
\end{eqnarray}
where the reduced (and projected) Green's function defined by
$\langle{\cal G}^{R,k}_{M}\rangle_\Lambda :=  \Lambda\langle{\cal
G}^{R,k}_{M}\rangle\overline\Lambda$ obeys
\begin{equation}
\langle{\cal G}^{R,k}_{M}\rangle_\Lambda
=
\langle{G_0}_{M}\rangle
- \textrm{i}\;\langle{G_0}_{M}\rangle\;
\sum_{i=k+1}^\infty {V^{\rm eff}_{M}}^{(i)}\;
\langle{\cal G}^{R,k}_{M}\rangle_\Lambda
\end{equation}
The crucial point is now that the kernel appearing in this integral
equation is obviously at least of ($k+1$)th order of the residual
kernel $V^{(3)}_R + \overline K_M^{(2)}$.  In particular the Neumann
series of $\langle{\cal G}^{R,k}_{M}\rangle_\Lambda$ consists, apart
from the 0th order term $\langle{G_0}_{M}\rangle$, only of terms of
order $> k$.  In other words, if we expand the propagator ${\cal
G}^{R,k}_{M}$, similar to the effective kernel ${V^{\rm eff}_{M}}$, in powers of the residual kernel $V^{(3)}_R + \overline
K_M^{(2)}$, {\it i.e.}
\begin{equation}
\label{GRk_expansion}
{\cal G}^{R,k}_{M} = \sum_{i=0}^\infty {{\cal G}^{R,k}_{M}}^{(i)}
\end{equation} 
and consider only terms up to $k$th order, we find
\begin{equation}
\left\langle \sum_{i=0}^k {{\cal G}^{R,k}_{M}}^{(i)}\right\rangle_{\!\!\Lambda} 
:=
\Lambda\left\langle \sum_{i=0}^k {{\cal G}^{R,k}_{M}}^{(i)}\right\rangle\overline\Lambda
=
\langle{G_0}_{M}\rangle.
\end{equation}  
This result now allows to state an appropriate approximation of the full Bethe-Salpeter equation
which is consistent with the Salpeter equation (\ref{SEquation_approx}):
replacing in the exact Bethe-Salpeter equation (\ref{BSEquation4}) the propagator ${\cal G}^{R,k}_{M}$
by its expansion (\ref{GRk_expansion}) up to the order $k$,
\begin{equation}
{\cal G}^{R,k}_{M} \longrightarrow \sum_{i=0}^k {{\cal G}^{R,k}_{M}}^{(i)}
\end{equation}
we obtain an approximation of the Salpeter amplitude
$\chi_M\simeq\chi_M^{(k)}$ by the solution of the approximated Bethe-Salpeter equation
\begin{eqnarray}
\label{BSEquation_approx}
\chi^{(k)}_{M} &=&
-\textrm{i}\;\sum_{i=0}^k {{\cal G}^{R,k}_{M}}^{(i)}\;\left(V_\Lambda^{(3)} + \sum_{i=1}^k {V^{\rm eff}_{M}}^{(i)}\right)\;\chi^{(k)}_{M}.
\end{eqnarray}
Then, the corresponding reduced amplitude
\begin{eqnarray}
\label{def:salpeter_ampl_CMS_approx}
\lefteqn{{\Phi_M^\Lambda}^{\!\!(k)} ({\bf p_\xi},{\bf p_\eta}) :=}\nn
&&\Lambda({\bf p_\xi},{\bf p_\eta})\;\int\frac{\textrm{d} p_\xi^0}{2\pi}\;\frac{\textrm{d} p_\eta^0}{2\pi}\;
\chi^{(k)}_{M}\left((p_\xi^0, {\bf p_\xi}), (p_\eta^0, {\bf p_\eta})\right),
\end{eqnarray}
is indeed the solution of the approximated Salpeter equation (\ref{SEquation_approx})
and $\chi_M^{(k)}$ can be reconstructed from ${\Phi_M^\Lambda}^{\!\!(k)}$
according to 
\begin{eqnarray}
\label{BSA_approx}
\chi_{M}^{(k)} &=&
-\textrm{i}\;\sum_{i=0}^k {{\cal G}^{R,k}_{M}}^{(i)}\left(V_\Lambda^{(3)} + \sum_{i=1}^k {V^{\rm eff}_{M}}^{(i)}\right)\;{\Phi^\Lambda_{M}}^{\!\!(k)}\!.
\end{eqnarray}
Our explicit model calculations so far have been performed in lowest
order, {\it i.e. }
with $V^{\rm eff}_{M}\simeq {V^{\rm eff}_{M}}^{(1)}$ and $\Phi^\Lambda_M \simeq{\Phi^\Lambda_M}^{\!\!(1)}$ 
as given by eqs. (\ref{Veff_Born_explicit}) and (\ref{Salp_Hamilt_V3_V2_born}). 
In this case we find from eq. (\ref{GRk})
\begin{eqnarray}
{\cal G}^{R,1}_{M}&\simeq& {{\cal G}^{R,1}_{M}}^{(0)} + {{\cal G}^{R,1}_{M}}^{(1)}\\
&=&
{G_0}_{M} 
- 
\textrm{i}\;{G_0}_{M}\;\left(\overline K^{(2)}_{M} +  V_R^{(3)} - {V^{\rm eff}_{M}}^{(1)}\right)\;{G_0}_{M}\nonumber
\end{eqnarray} 
such that the Bethe-Salpeter amplitude $\chi_M\simeq\chi_M^{(1)}$ in the corresponding order of approximation
is determined by
\begin{eqnarray}
\label{BSA_approxBorn}
\chi_{M}^{(1)} &=&
\left[{G_0}_{M} 
- 
\textrm{i}\;{G_0}_{M}\;\left(\overline K^{(2)}_{M} +  V_R^{(3)} - {V^{\rm eff}_{M}}^{(1)}\right)\;{G_0}_{M}
\right]\nn
&&\quad\qquad\times\;(-\textrm{i})\;\left[V_\Lambda^{(3)} + {V^{\rm eff}_{M}}^{(1)}\right]\;{\Phi^\Lambda_{M}}^{(1)}.
\end{eqnarray}}


\end{document}